\documentclass[aps,superscriptaddress,notitlepage,reprint]{revtex4-2}
\usepackage{epsfig,color}
\usepackage{graphicx}
\usepackage{dcolumn}
\usepackage{bm}
\usepackage{amsmath,amsfonts,amssymb,mathrsfs}
\usepackage{pstricks}
\usepackage{amsxtra}
\usepackage{cases}
\usepackage{amsthm}
\usepackage{braket}
\usepackage{natbib}
\usepackage{physics}
\usepackage{hyperref,color}
\usepackage{lipsum}
\definecolor{darkblue}{rgb}{0.0,0.0,0.7}
\hypersetup{colorlinks,breaklinks,linkcolor=darkblue,urlcolor=darkblue,anchorcolor=darkblue,citecolor=darkblue}

\providecommand{\ah}{{\hat{a}}}
\providecommand{\ad}{{\hat{a}^\dagger}}
\providecommand{\sh}{{\hat{s}}}
\renewcommand{\sd}{{\hat{s}^\dagger}}
\providecommand{\shz}{{\hat{\sigma}_z}}

\providecommand{\shp}{{\hat{\sigma}_+}}
\providecommand{\shm}{{\hat{\sigma}_-}}
\providecommand{\bh}{{\hat{b}}}
\providecommand{\bd}{{\hat{b}^\dagger}}
\providecommand{\Hh}{{\hat{H}}}
\providecommand{\rh}{{\hat{\rho}}}
\providecommand{\xh}{{\hat{\chi}}}

\providecommand{\rt}{\Tilde{\rho}}

\providecommand{\Jh}{\hat{J}}

\makeatletter
\def\@fnsymbol#1{\ensuremath{\ifcase#1\or \dagger\or *\or \ddagger\or
\mathsection\or \mathparagraph\or \|\or **\or \dagger\dagger\or \ddagger\ddagger \else\@ctrerr\fi}}
\makeatother

\begin{document}

\title{Engineering squeezed thermal reservoirs via passive linear coupling}

\author{Cheng-Lin Lee}
\thanks{These authors contributed equally to this work.\label{equal}}
\affiliation{Department of Physics and Center for Theoretical Physics, National Taiwan University, Taipei 106319, Taiwan}

\author{Chiao-Hsuan Wang{$^{\ref{equal},}$}}
\email{chiaowang@phys.ntu.edu.tw}
\affiliation{Department of Physics and Center for Theoretical Physics, National Taiwan University, Taipei 106319, Taiwan}
\affiliation{Center for Quantum Science and Engineering, National Taiwan University, Taipei 106319, Taiwan}
\affiliation{Physics Division, National Center for Theoretical Sciences, Taipei 106319, Taiwan}

\begin{abstract}
Squeezed thermal reservoirs, characterized by thermal noise with anisotropic fluctuations, have profound implications in quantum thermodynamics and serve as powerful resources for quantum information. However, their experimental realizations remain challenging.  Existing schemes typically rely on injected squeezed light, time-dependent modulation, or driven nonlinear interactions, which introduce complexity and limit experimental feasibility. Using only time-independent linear coupling to a lossy mode within a normal thermal environment, we identify a general and experimentally accessible framework for squeezed-reservoir engineering, applicable across platforms such as circuit and cavity quantum electrodynamics as well as coupled cavity systems. We illustrate the framework through two experimentally relevant cases: directional phase coherence extension in two-level systems like qubits or atoms, and dissipative quadrature squeezing in bosonic modes like photons or phonons.  By eliminating the need for active control or squeezed input, our passive linear-coupling approach provides a resource-efficient and practical pathway to dissipative squeezing, decoherence suppression, entanglement stabilization, quantum simulation, and the exploration of unconventional quantum thermodynamics and phase transitions.
\end{abstract}

\maketitle

\section{Introduction}
Quantum systems are inherently susceptible to environmental influences, leading to dissipative effects that induce decoherence and diminish quantum properties. Quantum reservoir engineering seeks to convert typically undesirable interactions between the quantum system and its environment into advantageous resources~\cite{Poyatos1996}. This powerful approach has broad applications, including quantum computation and state preparation~\cite{Verstraete2009,Carvalho2001}, entanglement generation~\cite{Krauter2011,Lin2013,Shankar2013}, quantum memory~\cite{Pastawski2011}, quantum metrology~\cite{Goldstein2011,Reiter2017}, quantum simulation~\cite{Kapit2014,Wang2019}, quantum error correction~\cite{Puri2019, Gertler2021}, quantum battery~\cite{Barra2019}, and quantum state transfer~\cite{Wang2019a}. Controlled coupling between the system and the reservoir may include additional effects such as non-Markovian behavior~\cite{Liu2011}, nonreciprocal transmission~\cite{Metelmann2015}, and heat engines with atypical thermodynamic properties~\cite{Abah2014,DeAssis2019}.

Squeezed thermal reservoirs are particularly unconventional and intriguing ~\cite{Gardiner1985,Gardiner2004,Manzano2018}. Squeezing, a quantum phenomenon originating from anomalous correlations that reduce fluctuations in one phase-space quadrature while increasing them in the conjugate quadrature ~\cite{Loudon1987,Kitagawa1993}, serves as a powerful resource in quantum information science ~\cite{Braunstein2005,Madsen2022} and has profound implications across various fields of physics. In quantum science and precision measurement, squeezed light is crucial for achieving precise measurements at the quantum limit in gravitational wave detection ~\cite{Bondurant1984,Jia2024}. In condensed matter physics, squeezing can induce phase transitions and enable the formation of exotic states of matter~\cite{Peano2016,Zhu2020}. In thermodynamics, quantum heat engines utilizing squeezed reservoirs can surpass classical efficiency limits ~\cite{Huang2012,Roßnagel2014,Manzano2016,Klaers2017}.
Additionally, quantum systems coupled to squeezed reservoirs can exhibit compelling behaviors, such as extended atomic lifetimes~\cite{Gardiner1986,Murch2013} and anomalous resonance fluorescence~\cite{Swain1994,Toyli2016}. Squeezed reservoirs also offer unique opportunities for generating dissipative squeezing ~\cite{Parkins2006,Porras2012,Kronwald2014,Wollman2015,Bai2021} and stabilizing entanglement ~\cite{Kraus2004,Banerjee2010,Wang2013,Zippilli2015,Govia2022}.

Despite their extensive applications, squeezed thermal reservoirs are challenging to realize, as they require anomalous correlations that are fundamentally absent in standard thermal environments. Typically, proposed schemes involve generating broadband squeezed light through externally driven nonlinear interactions~\cite{Murch2013} or utilizing driven-dissipative dynamics in platforms like trapped ions~\cite{Poyatos1996} and optomechanical systems~\cite{Kronwald2013,Wollman2015}. Alternatively, squeezed dissipation can be engineered via time-dependent modulation of system parameters~\cite{Shahmoon2013,Govia2022}. These approaches necessitate additional experimental overhead of nonlinear elements and active external drives, which complicates their implementation.

While few experimental milestones have been achieved~\cite{Murch2013,Wollman2015}, the broader realization of squeezed thermal reservoirs remains limited by the reliance on active control and nonlinear resources, which may hinder integration across diverse experimental platforms. In this work, we address this challenge through a physical mechanism for squeezed reservoir engineering based on the interference between distinct system-bath interaction processes enabled by a passive, linear coupling framework. In contrast to previous active schemes, our approach utilizes only static, bilinear interactions ubiquitous in many quantum systems. This resource-efficient design eliminates the need for active drives or squeezed-light injection, providing a practical and experimentally accessible pathway to squeezed reservoir realization.

The remainder of this article is organized as follows. In Sec.~II, we introduce a general framework for squeezed reservoir engineering based on passive bilinear system-bath coupling. In Sec.~III, we quantify the reservoir properties by establishing the dependence of squeezing parameters on the underlying physical configurations. To showcase the versatility and experimental relevance of our approach, we apply this framework to two representative cases: a two-level system (Sec.~IV), representing qubits and atoms in circuit and cavity quantum electrodynamics, and a bosonic mode (Sec.~V), applicable to coupled photonic and phononic systems. Finally, we discuss the implications for advancing quantum technology in Sec.~VI.

\section{Overview of the squeezed reservoir engineering framework}
We begin by outlining the physical mechanism underlying our squeezed reservoir engineering framework, which arises from the interference between distinct interaction processes enabled by passive, time-independent linear interactions.
This passive configuration effectively induces squeezed dissipation in the target system through coupling to a lossy bosonic mode embedded in a normal thermal environment, without requiring squeezed input or dynamic modulation. The following generic model captures the core mechanism of reservoir-induced squeezing and serves as the foundation for the two representative examples presented in this work.

We consider the bilinear coupling between a system mode $S$ and a dissipative bosonic mode $B$, referred to as the bath mode, described by the total Hamiltonian
\begin{align}
\Hh_{\rm tot}= & \, \, \hbar \omega_s \sd \sh +\hbar \omega_b \bd \bh
\notag\\
&+ \hbar g (\sh e^{-i \varphi_s} + \sd e^{i \varphi_s}) (\bh e^{-i \varphi_b}+ \bd e^{i \varphi_b}),
\label{eqn:Htot}
\end{align}
where $\sd/\sh$ are the creation/annihilation operators of the system mode of frequency $\omega_s$, $\bd/\bh$ are the creation/annihilation operators of the bath mode of frequency $\omega_b$, and $g$ is the coupling strength between the system and the bath modes.  We assume $g$ to be real and positive without loss of generality by introducing complex phases $\varphi_s$ and $\varphi_b$ for the mode operators.

\begin{figure}[htbp]
\begin{center}
\includegraphics[width= \linewidth]{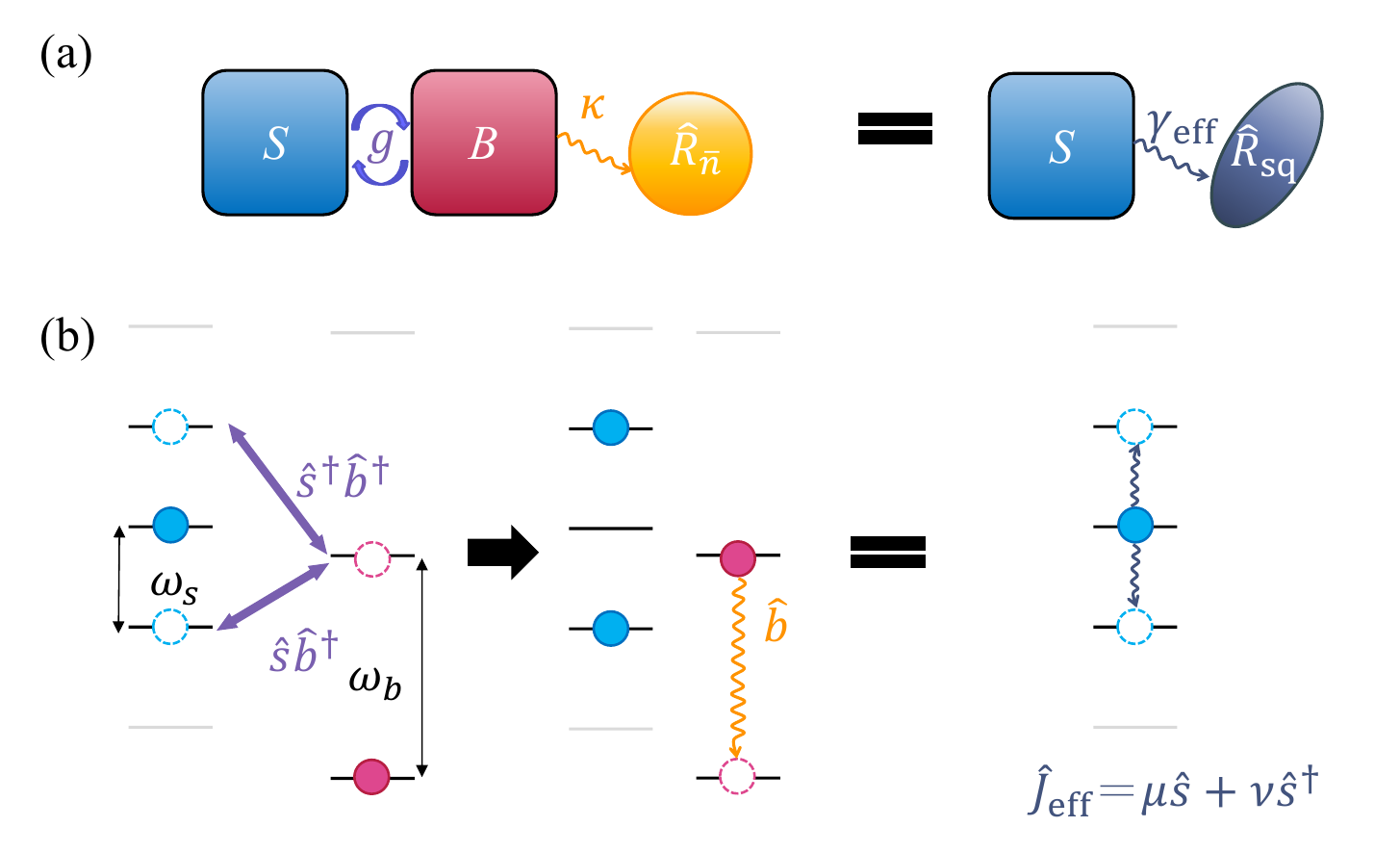}
\caption{Schematics of the squeezed thermal reservoir engineering method. (a) A quantum system ($S$) is linearly coupled to a dissipative bosonic mode (bath mode $B$) with strength $g$, which in turn dissipates into a thermal environment $\hat{R}_{\bar{n}}$ at rate $\kappa$. In the off-resonant coupling or bad-cavity regime, this setup effectively realizes a squeezed thermal reservoir $\hat{R}_{\rm sq}$ for $S$, with an effective dissipation rate $\gamma_{\rm eff}$.
(b) Diagrammatic representation of the second-order processes resulting in the effective squeezed jump operator $\hat{J}_{\rm eff}$. Narrow and wide ladders indicate the energy levels of the system (frequency $\omega_s$) and the bath (frequency $\omega_b$), respectively. Solid circles mark initial states, and dashed circles mark final states after transitions. Number-conserving ($\hat{s}^\dagger \hat{b}$) and nonconserving ($\hat{s} \hat{b}$) transitions interfere, giving rise to the effective squeezed dissipation illustrated in the right panel.}
\label{fig:schematic} 
\end{center}
\end{figure}

In the configuration presented in Fig.~\ref{fig:schematic}(a), the system mode interacts only with the bath mode, while the bath mode is also coupled to a thermal environment. The evolution of the combined $S{+}B$ modes is governed by the Lindblad master equation~\cite{Breuer2002},
\begin{align}
    \dot{\xh}=-\frac{i}{\hbar}[\Hh_{\rm tot},\xh]+(\bar{n}+1) \kappa \mathcal{D}[\bh] (\xh)+\bar{n}\kappa \mathcal{D}[\bd] (\xh) ,
    \label{eqn:totalmaster}
\end{align}
where $\xh$ is the total density matrix of the combined $S+B$ modes, $\kappa$ is the dissipation rate of the bath mode, and $\mathcal{D}[\hat{J}] (\xh)\equiv\hat{J} \xh \hat{J}^{\dagger}-\frac{1}{2}\hat{J}^{\dagger} \hat{J} \xh -\frac{1}{2}\xh\hat{J}^{\dagger} \hat{J}$ is the Lindblad dissipator describing the dissipative dynamics associated with a jump operator $\hat{J}$. The thermal environment is modeled as a multimode thermal state $\hat{R}_{ \bar{n}}$ with a mean thermal occupation number $\bar{n}$ at the bath mode frequency $\omega_b$. This gives rise to a relaxation rate of $\left(\bar{n}+1\right)\kappa$, associated with the jump operator $\bh$, and a heating rate of $\bar{n}\kappa$, associated with the jump operator $\bd$, for the bath mode.

In the off-resonant coupling regime, $\abs{\omega_b-\omega_s}  \gg g$, or assuming a bad cavity, $\kappa \gg g$, the thermal environment dissipates the system mode through second-order processes via standard adiabatic elimination of the bath mode~\cite{Gardiner2004}.  Consider the process shown in Fig.~\ref{fig:schematic}(b), where dissipation arises from coherent excitation of the bath mode followed by its relaxation into the thermal environment. The system-bath interaction includes two relevant processes, both accompanied by bath mode excitation: the number-conserving term $\sh \bd$, which causes system mode deexcitation, and the number-nonconserving term $\sd \bd$, which causes system mode excitation. The bath mode then resets through relaxation into its thermal environment, governed by the jump operator $\bh$. This overall process gives rise to an effective jump operator for the system mode, $\hat{J}_{\rm eff}=\mu \sh+\nu \sd$, representing a superposition of excitation and deexcitation transitions within the system mode. 

A squeezed relaxation naturally emerges from the combined effect of the number-conserving and number-nonconserving processes.
Because the two processes involve different energy differences, $\Delta E = \hbar(\omega_b \mp \omega_s)$, they occur at distinct transition rates and result in unequal jump operator coefficients, $|\mu| \neq |\nu|$.
The squeezing effect can be identified by appropriately normalizing the effective jump operator, which then takes the form of a squeezed system operator~\cite{Gardiner1985},
\begin{align}
    \hat{J}_{\rm eff}=\hat{\mathcal{S}}^{\dagger}_{\xi}\sh \hat{\mathcal{S}}_{\xi}\equiv\cosh(r) \sh - e^{i \theta}\sinh(r) \sd,
\end{align}
where $\xi = r e^{i\theta}$ is the complex squeezing parameter, with squeezing strength $r \geq 0$ and squeezing angle $\theta \in [0, 2\pi)$, defined throughout this work. This structure indicates that dissipation due to $\hat{J}_{\rm eff}$ corresponds to relaxation of the squeezed system mode.

Squeezed heating of the system mode also emerges from the complementary mechanism, if the environment coupled to the bath mode is at finite temperature. The heating process is governed by the adjoint of the relaxation jump operator $\hat{J}_{\rm eff}^{\dagger}=\hat{\mathcal{S}}^{\dagger}_{\xi}\sd \hat{\mathcal{S}}_{\xi}$, which can be interpreted as heating of the squeezed system mode.  Including both bath mode relaxation and heating processes, the resulting dissipative dynamics of the system is governed by an effective superoperator in the Lindbladian form:
\begin{align}
    \mathcal{L}^{D}_{\rm eff}(\hat{\rho}_s)=(\bar{n}+1)\gamma_{\rm eff}\mathcal{D}[\hat{\mathcal{S}}^{\dagger}_{\xi}\sh \hat{\mathcal{S}}_{\xi}](\rh_{s})+\bar{n}\gamma_{\rm eff}\mathcal{D}[\hat{\mathcal{S}}^{\dagger}_{\xi}\sd \hat{\mathcal{S}}_{\xi}](\rh_{s}),
    \label{eqn:effL}
\end{align}
where $\hat{\rho}_s = \text{Tr}_{B}({\hat{\chi}})$ is the reduced density matrix of the system mode after tracing out the bath mode. 

Substituting the squeezing structure of $\hat{J}_{\rm eff}$ into the Lindblad dissipator, Eq.~\eqref{eqn:effL} becomes
\begin{align}
\mathcal{L}^{D}_{\rm eff}(\rh_s) =& (N+1)\frac{\gamma_{\rm eff}}{2}\left(2\hat{s}\rh_s \hat{s}^\dagger - \hat{s}^\dagger \hat{s}\rh_s - \rh_s \hat{s}^\dagger \hat{s}\right)\notag \\
&+ N\frac{\gamma_{\rm eff}}{2}\left(2\hat{s}^\dagger \rh_s \hat{s} - \hat{s} \hat{s}^\dagger \rh_s - \rh_s \hat{s} \hat{s}^\dagger\right) \notag \\
&- M\frac{\gamma_{\rm eff}}{2}\left(2\hat{s}^\dagger \rh_s \hat{s}^\dagger - \hat{s}^\dagger \hat{s}^\dagger \rh_s - \rh_s \hat{s}^\dagger \hat{s}^\dagger \right)
\notag \\
&- M^{*}\frac{\gamma_{\rm eff}}{2}\left(2\hat{s} \rh_s \hat{s} - \hat{s} \hat{s} \rh_s - \rh_s \hat{s} \hat{s} \right),
\label{eqn:Lsq}
\end{align}
where $N = \bar{n}\cosh^2(r) + (\bar{n}+1)\sinh^2(r)$ corresponds to a modified thermal occupation number, and $M = (2\bar{n}+1)\cosh(r)\sinh(r) e^{i\theta}$ represents the squeezing-induced two-excitation correlation terms, absent in normal thermal environments. These two-excitation correlations cannot be produced by simply rescaling temperature, modifying coupling strength, or altering the reservoir spectral density of a normal thermal reservoir.

The effective dissipation is formally equivalent to that of a system coupled to a broadband squeezed thermal reservoir~\cite{Breuer2002}.  This engineered reservoir is characterized by a multi-mode squeezed thermal state $\hat{R}_{\rm sq}=\bigotimes_j\hat{\mathcal{S}}_{j,\xi}^\dag\hat{R}_{\bar{n}_j}\hat{\mathcal{S}}_{j,\xi}$, where $\hat{R}_{\bar{n}_j}$ is the thermal state of a reservoir mode $j$ with mean thermal occupation number $\bar{n}_j$. In this squeezed thermal reservoir, fluctuations of the phase-space quadratures are exponentially suppressed (squeezed) by a factor of $e^{-r}$ along the angle $ \frac{\theta}{2} + \frac{\pi}{2} $ from the real axis in phase space, and enhanced (anti-squeezed) by a factor of $e^{r}$ along the conjugate angle $ \frac{\theta}{2} $. 
Such quadrature-dependent fluctuations are a characteristic feature of squeezed reservoirs. They originate from the two-excitation correlations ($M, M^{*}$) absent in normal thermal environments, giving rise to unconventional system dynamics.

In prior realizations, these two-excitation correlations are typically generated through driven nonlinear interactions or active parametric processes that inject externally produced squeezed noise into the system~\cite{Murch2013,Wollman2015}. In contrast, in the present framework the same effective correlated dissipator emerges from the interference between number-conserving and number-nonconserving interaction processes mediated by a dissipative auxiliary mode. The two-excitation correlations therefore emerge intrinsically from time-independent bilinear coupling, providing a passive and linear mechanism for squeezed dissipation without dynamic modulation or nonlinear hardware.

This interference-based mechanism applies broadly to systems coupling linearly with a dissipative bosonic mode in the off-resonant or bad-cavity regime. The resulting correlations underlie the unconventional dynamics observed in squeezed reservoirs, including extended qubit lifetimes and anomalous resonance fluorescence~\cite{Gardiner1986,Murch2013,Swain1994,Toyli2016}, steady-state entanglement generation~\cite{Kraus2004,Banerjee2010,Wang2013,Zippilli2015,Govia2022}, thermodynamic advantages in quantum heat engines~\cite{Huang2012,Roßnagel2014,Manzano2016,Klaers2017}, and dissipative squeezing for sensing applications~\cite{Parkins2006,Porras2012,Kronwald2014,Wollman2015,Bai2021}.
As bilinear couplings are ubiquitous in platforms including circuit and cavity quantum electrodynamics, and coupled photonic or phononic systems, the present mechanism provides a general and resource-efficient route to squeezed reservoir engineering.

\section{Squeezing parameter of the engineered reservoir \label{sec:sqpara}}
To quantify the engineered squeezing, we extract the squeezing parameter $\xi$ and the effective dissipation rate $\gamma_{\rm eff}$ experienced by the system mode, which depend on the microscopic parameters of the entire system-bath-environment model $(\omega_s, \omega_b, g, \kappa, \bar{n}, \varphi_s)$. Specifically, they can be derived from the second-order transition amplitudes, leading to the expression
\begin{align}
\sqrt{\gamma_{\rm eff}}\hat{J}_{\rm eff}&=\sqrt{\gamma_{\rm eff}}[\cosh(r) \sh - e^{i \theta}\sinh(r) \sd]\notag\\
&=e^{i \varphi_j}\left( \frac{-g\sqrt{\kappa}e^{-i\varphi_s}}{\omega_b- \omega_s -i \kappa/2}\sh+\frac{-g\sqrt{\kappa}e^{i\varphi_s}}{\omega_b + \omega_s -i \kappa/2}\sd \right),
\label{eqn:Jeff}
\end{align}
where $e^{i \varphi_j}$, $\varphi_j = \pi +\varphi_s - \tan^{-1}\left[\frac{\kappa}{2(\omega_b - \omega_s)}\right]$, is a global phase that does not affect the Lindblad dissipator. The two terms following the last equality correspond to number-conserving and number-nonconserving processes, each broadened by the dissipation rate $\kappa$. 

Extracting from the form of the squeezed jump operator [Eq.~\eqref{eqn:Jeff}], the squeezing parameter $\xi=r e^{i \theta}$ and the effective dissipation rate $\gamma_{\rm eff}$ are given by
\begin{align}
r &= \tanh^{-1}\left(\sqrt{\frac{(\omega_b - \omega_s)^2 + \kappa^2/4}{(\omega_b + \omega_s)^2 + \kappa^2/4}}\right), \notag\\
\theta &= 2\varphi_s + \pi + \arg\left(\frac{\omega_b - \omega_s - i\kappa/2}{\omega_b + \omega_s - i\kappa/2}\right), \notag\\
\gamma_{\rm eff} &= \frac{4g^2 \kappa \omega_b \omega_s}{\left[(\omega_b - \omega_s)^2 + \kappa^2/4\right]\left[(\omega_b + \omega_s)^2 + \kappa^2/4\right]}.
\label{eqn:sqparam}
\end{align}
The dissipative contribution to the system dynamics is fully governed by the effective squeezed Lindbladian characterized by the above parameters.
The formal derivation of the effective master equation that includes Hamiltonian corrections~\cite{Jager2022,Jager2023}, consistent with the second-order transition interference picture described above, is provided in Appendix~\ref{AppendixA}.

\begin{figure}[htbp]
\begin{center}
\includegraphics[width=\linewidth]{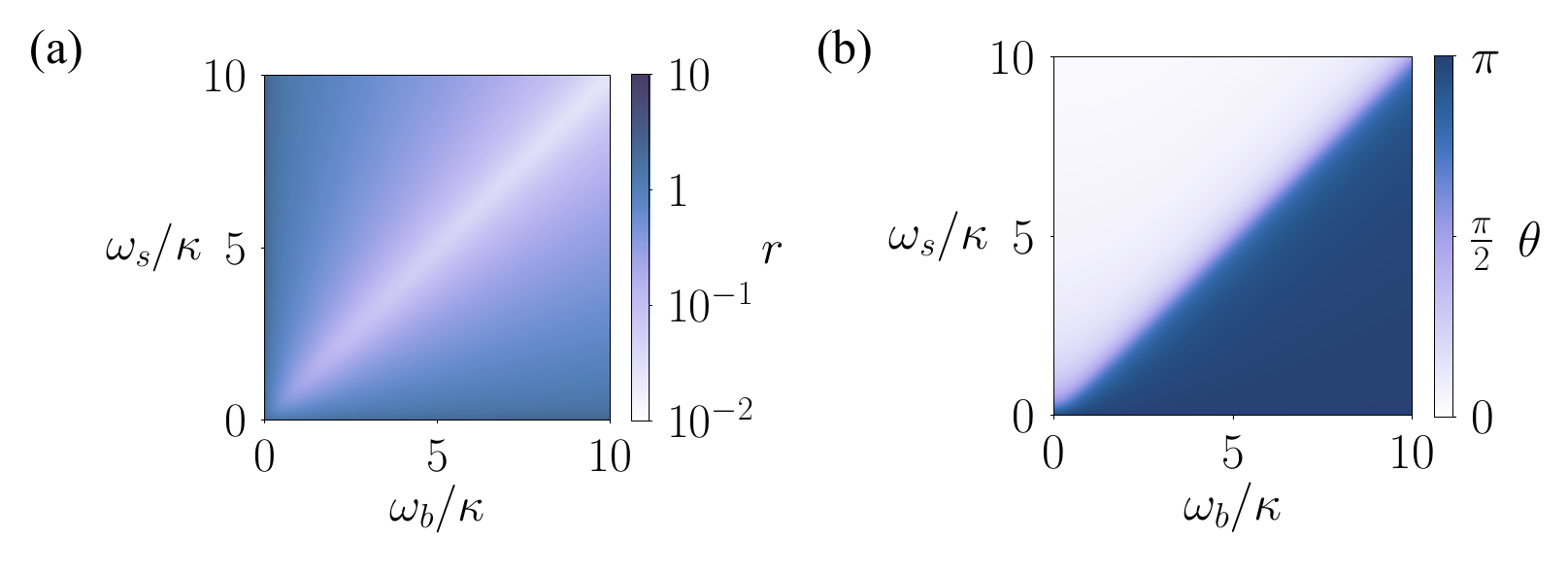}
\caption{Squeezing properties of the engineered reservoir. (a) Squeezing strength $r$ as a function of the dimensionless mode frequencies $\omega_s/\kappa$ and $\omega_b/\kappa$, clearly revealing enhanced squeezing at larger detunings. (b) Squeezing angle $\theta$ as a function of the dimensionless mode frequencies $\omega_s/\kappa$ and $\omega_b/\kappa$. Both plots are evaluated with 
$\varphi_s = 0$, showing how the reservoir squeezing properties depend on the system-bath frequency configuration.
}
\label{fig:sqparameter} 
\end{center}
\end{figure}

To illustrate how the engineered squeezing can be tailored through microscopic parameters,  we plot the squeezing strength $r$ and the squeezing angle $\theta$ as functions of the dimensionless mode frequencies $\omega_s/\kappa$ and $\omega_b/\kappa$ in Fig.~\ref{fig:sqparameter}. Near resonance, $\omega_s \approx \omega_b$, number-conserving contributions dominate, and the reservoir behaves like a normal thermal environment with minimal squeezing. As the detuning $|\omega_s - \omega_b|$ increases, number-nonconserving contributions become more prominent, leading to stronger squeezing. This trend is indicated by Fig.~\ref{fig:sqparameter}(a) through the monotonic increase of $r$ with detuning. The corresponding squeezing angle $\theta$ is shown in Fig.~\ref{fig:sqparameter}(b).

We have established that the dissipative dynamics of a quantum system, bilinearly coupled to a lossy bosonic mode in the off-resonant or bad-cavity regime, can effectively emulate those of a system interacting with a squeezed thermal reservoir, as in Fig.~\ref{fig:schematic}(a). The resulting engineered dissipation provides a general and resource-efficient framework for realizing squeezed reservoirs using only passive linear interactions. To illustrate the experimental feasibility and versatility of this method, we now investigate two representative examples for the system mode: two-level systems, such as qubits or atoms (quantum Rabi model), and bosonic systems, such as photons or phonons (bosonic interaction model). Both models highlight how such engineered dissipation governs quantum dynamics in physically relevant regimes.

\section{Qubit dynamics in an engineered squeezed reservoir}
\begin{figure}[htbp]
\begin{center}
\includegraphics[width= \linewidth]{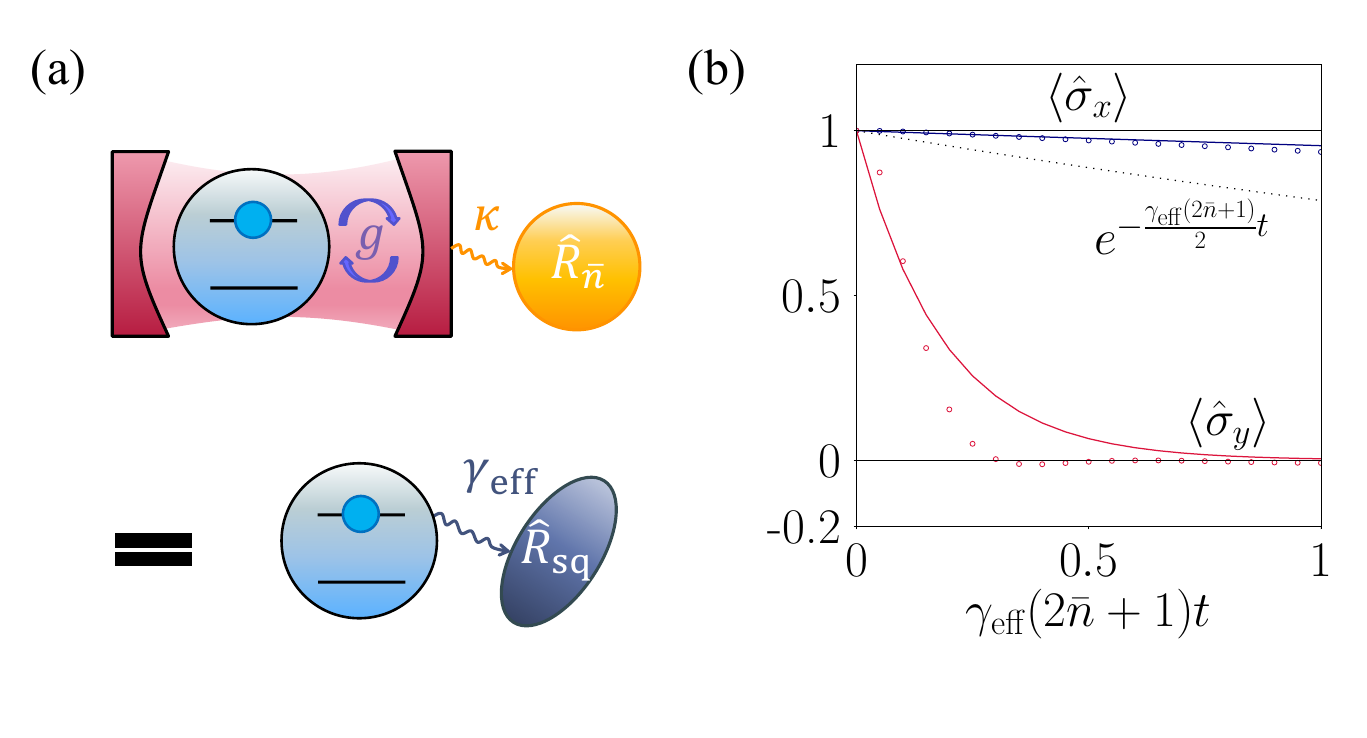}
\caption{Qubit dynamics under an engineered squeezed thermal reservoir. (a) Schematic of the quantum Rabi model implementation of the squeezed reservoir engineering framework, where a qubit is coupled to a lossy cavity dissipating into a thermal environment. This configuration effectively realizes a squeezed thermal reservoir for the qubit.
(b) Simulated qubit dynamics showing the evolution of $\langle \hat{\sigma}_x \rangle$ and $\langle \hat{\sigma}_y \rangle$, starting from initial states aligned along the $x$ and $y$ axes, respectively. Solid lines represent the effective master equation~(\ref{eqn:qubitmaster}), and circles indicate results from the full model~(\ref{eqn:totalmaster}); the dotted line shows decay under a normal thermal reservoir for comparison. Simulations are performed with parameters relevant to fluxonium-based circuit QED devices~\cite{Manucharyan2009,Nguyen2019,Nguyen2022}: $\omega_q / 2 \pi = 100$ MHz, $\omega_b / 2 \pi = 1.0$ GHz, $\kappa / 2 \pi = 600$ MHz, $g / 2 \pi = 176$ MHz, and $\bar{n} = 10$, yielding a reservoir squeezing parameter $\xi \approx -1.2$.
}
\label{fig:QRM}
\end{center}
\end{figure}

As a representative implementation of our squeezed reservoir engineering framework, we consider a qubit, modeled as a two-level system, bilinearly coupled to a dissipative bosonic mode. This architecture is widely realized in a variety of quantum platforms, including cavity quantum electrodynamics (cavity QED) with neutral atoms~\cite{Haroche1989,Raimond2001}, circuit quantum electrodynamics (circuit QED) with superconducting qubits~\cite{Blais2021a,Wallraff2004}, and trapped-ion systems~\cite{Leibfried2003,Lv2018}. A schematic of this setup is shown in Fig.~\ref{fig:QRM}(a), where the lossy cavity mode serves as the engineered squeezed reservoir for the qubit.

In this setup, the total Hamiltonian is described by the quantum Rabi model,
\begin{align}
    \Hh_{\rm Rabi}=\frac{\hbar \omega_q}{2} \shz +\hbar \omega_b \bd \bh + \hbar g (\shm + \shp) (\bh + \bd),
\end{align}
where we set $s=q$, $ \sd=\shp$, $ \sh=\shm$, and take $\varphi_q=\varphi_b=0$.
The quantum Rabi model describes the bilinear coupling between a two-level system and a single bosonic mode~\cite{Rabi1936,Rabi1937}, and provides a unified framework for modeling light-matter interaction across these diverse experimental platforms.

In the bad-cavity or off-resonant regime, the lossy cavity mode can be adiabatically eliminated, yielding an effective description of the qubit coupled to a squeezed thermal reservoir. The resulting qubit Hamiltonian takes the form (see Appendix~\ref{sec:Hqeff})
\begin{align}
    \Hh_{q,\rm eff}&=\frac{\hbar \omega_{q,\rm eff}}{2}\shz.
\end{align}
The dynamics of the qubit is governed by the effective master equation
\begin{align}
    \dot{\rh}_q=&-\frac{i}{\hbar}[\Hh_{q,\rm eff},\rh_q]+(\bar{n}+1)\gamma_{\rm eff}\mathcal{D}[\shm^{\xi}](\rh_{q})\notag\\
    &+\bar{n}\gamma_{\rm eff}\mathcal{D}[\shp^{\xi}](\rh_{q}),
\label{eqn:qubitmaster}
\end{align}
where $\hat{\sigma}_{\pm}^{\xi}=\cosh(r) \hat{\sigma}_{\pm} - e^{\mp i \theta}\sinh(r) \hat{\sigma}_{\mp}$ and $\rh_{q}$ is the density matrix of the qubit.
This master equation captures the dynamics of a qubit interacting with a squeezed thermal reservoir, as shown in the lower panel of Fig.~\ref{fig:QRM}(a).

It has been theoretically investigated and experimentally demonstrated that a squeezed thermal reservoir can suppress phase decay in qubits, leading to extended qubit lifetimes~\cite{Gardiner1986,Murch2013}. Consider the case where $\omega_{q,\rm eff}=0$ and $\xi =-r$; the qubit dynamics in a squeezed thermal reservoir is described by~\cite{Gardiner1986,Gardiner2004}
\begin{align}
\left\langle \dot{\hat{\sigma}}_x\right\rangle&= -\frac{\gamma_{\rm eff}}{2}\left(2\bar{n}+1\right)e^{-2r}\left\langle \hat{\sigma}_x\right\rangle\equiv -\gamma_x \left\langle \hat{\sigma}_x\right\rangle,\notag\\
\left\langle \dot{\hat{\sigma}}_y\right\rangle&=-\frac{\gamma_{\rm eff}}{2}\left(2\bar{n}+1\right)e^{2r}\left\langle \hat{\sigma}_y\right\rangle\equiv -\gamma_y \left\langle \hat{\sigma}_y\right\rangle,\notag\\
\left\langle \dot{\hat{\sigma}}_z\right\rangle&=-\gamma_{\rm eff} \left(2\bar{n}+1\right)\cosh(2r)\left\langle \hat{\sigma}_z\right\rangle-\gamma_{\rm eff}\notag\\&\equiv -\gamma_z \left\langle \hat{\sigma}_z\right\rangle-\gamma_{\rm eff}.
\label{eqn:Bloch}
\end{align}

Compared to radiative decay in a normal thermal environment with a mean thermal occupation number $\bar{n}$ at rate $\gamma_{\rm eff}$, the relaxation rates satisfy $2\gamma_x=2\gamma_y=\gamma_z=\gamma_{\rm eff}(2\bar{n}+1)$, which defines the longitudinal lifetime $T_1=1/\gamma_z$ and an isotropic transverse lifetime $T_2=1/\gamma_x=1/\gamma_y=2T_1$. In contrast, the longitudinal relaxation rate $\gamma_z$ is enhanced by a factor of $\cosh(2r)$ in the squeezed reservoir, while the transverse relaxation rates $\gamma_x$ and $\gamma_y$ are suppressed and enhanced by factors of $e^{-2r}$ and $e^{2r}$, respectively, reflecting the reduction and amplification of fluctuations in the corresponding reservoir quadratures.

Due to these asymmetric fluctuations, the transverse coherence times become direction dependent: $T_{2x} = 1/\gamma_x$ can exceed $2T_1$, while $T_{2y} = 1/\gamma_y$ is correspondingly reduced. This departure from the thermal bound $T_2 \le 2T_1$ constitutes a distinctive signature of squeezed reservoirs~\cite{Gardiner1986} and serves as a key experimental indicator for verifying squeezed environments~\cite{Murch2013}.

This directional coherence enhancement cannot be reproduced by anisotropic relaxation into independent thermal reservoirs along the $x$, $y$, and $z$ axes with rates $\Gamma_x$, $\Gamma_y$, and $\Gamma_z$.  In such a scenario, the overall decay rate along the Bloch sphere direction $i$ takes the form $\gamma_i = \Gamma_i + (\Gamma_j + \Gamma_k)/2$ (for distinct $i,j,k \in \{x,y,z\}$), which enforces the constraint $\gamma_i \ge \gamma_j/2$. Consequently, the inequality $T_2 \le 2T_1$ remains satisfied for any combination of incoherent thermal processes, indicating that $T_{2x} > 2T_1$ requires the correlations intrinsic to the squeezed dissipator.

To examine the reservoir squeezing effect and its experimental relevance, we simulate the qubit dynamics using parameters motivated by fluxonium-based circuit QED systems~\cite{Manucharyan2009,Nguyen2019,Nguyen2022}, which support strong coupling between low-frequency qubits and microwave resonators.  Details of the numerical simulations can be found in Appendix~\ref{sec:qubitsim}. The simulations reveal a clear anisotropy in qubit lifetimes along the two transverse directions, as shown in Fig.~\ref{fig:QRM}(b), where the lifetime of $\langle \hat{\sigma}_x \rangle$ is significantly extended, while that of $\langle \hat{\sigma}_y \rangle$ is shortened, compared to the normal thermal decay. These results are consistent with the effective dynamics described by Eq.~\eqref{eqn:Bloch}, where the transverse relaxation rate $\gamma_x$ is suppressed while $\gamma_y$ is enhanced. This behavior reflects the anisotropic dissipation characteristic of squeezed reservoir engineering in our framework.

\section{Cavity dynamics in an engineered squeezed reservoir}
\begin{figure}[htbp]
\begin{center}
\includegraphics[width=\linewidth]{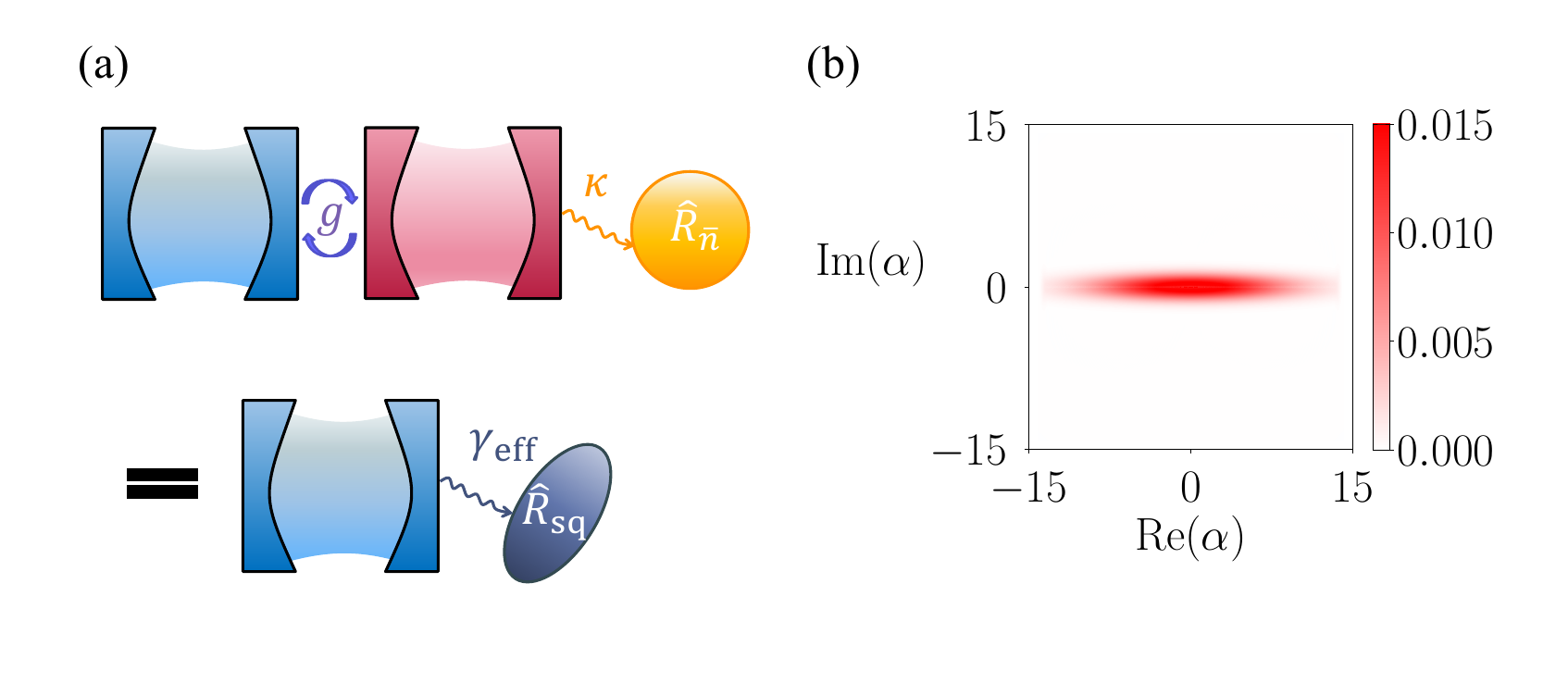}
\caption{Cavity dynamics under an engineered squeezed thermal reservoir. (a) Schematic of the cavity implementation of the squeezed reservoir engineering framework, where a system cavity (blue) is coupled to a lossy auxiliary cavity mode (red) that dissipates into a normal thermal environment. This configuration effectively realizes a squeezed thermal reservoir for the system cavity.
(b) Simulated steady-state Wigner function of the system cavity in the off-resonant and bad-cavity regime. The elliptical shape in phase space reflects quadrature-dependent fluctuations associated with engineered squeezed dissipation. Simulations are performed with parameters relevant to coupled microwave cavity systems~\cite{Krantz2019, Miyanaga2021}: $\omega_a / 2 \pi = 1.12$ GHz, $\omega_b / 2 \pi = 4.4$ GHz, $\kappa / 2 \pi = 4.0$ GHz, $g / 2 \pi = 1.2$ GHz, and $\bar{n} = 0$, yielding a steady-state squeezing parameter $\xi_a \approx 1$.}
\label{fig:BB} 
\end{center}
\end{figure}

For the second implementation of our squeezed reservoir engineering framework, we consider a bosonic system mode, such as a photonic, microwave, or acoustic cavity, bilinearly coupled to another dissipative bosonic mode. A schematic of this configuration is shown in Fig.~\ref{fig:BB}(a), where a lossy cavity mode serves as an engineered squeezed reservoir for the system cavity mode.

In the bosonic (cavity) implementation, we take the photonic cavity mode $a$ as the system and a lossy auxiliary photonic cavity mode $b$ as the bath. Setting $s = a$, $\sd = \ad$, $\sh = \ah$, and taking $\varphi_a = \varphi_b = 0$, the total Hamiltonian takes the form
\begin{align} \Hh_{\rm Bosonic} = \hbar \omega_a \ad \ah + \hbar \omega_b \bd \bh + \hbar g (\ah + \ad)(\bh + \bd), \end{align}
which describes a bilinear interaction between two bosonic modes. We refer to this setting as the bosonic interaction model. Such interactions are widely implemented across a variety of quantum platforms, including inductive couplings (with $\varphi_a = \varphi_b = 0$) or capacitive couplings (with $\varphi_a = \varphi_b = \pi/2$) between superconducting resonators~\cite{Krantz2019,Rasmussen2021}, as well as electromechanical interactions in piezoelectric devices~\cite{Zeuthen2018,Blesin2021}.

In the bad-cavity or off-resonant regime, the bath mode can be adiabatically eliminated, yielding an effective quadratic system Hamiltonian with a shifted frequency $\omega_{a,\rm eff}$ and a single-mode squeezing term of strength $\Lambda$ (see Appendix \ref{sec:Haeff}),
\begin{align}
    \Hh_{a, \rm eff}= \hbar \omega_{a,\rm eff} \ad \ah + \frac{i \hbar}{2} (\Lambda \hat{a}^{\dagger 2} -\Lambda^{*} \ah^2),
\end{align}
The dynamics of the system cavity is governed by the effective master equation
\begin{align}
    \dot{\rh}_a=&-\frac{i}{\hbar}[\Hh_{a,\rm eff},\rh_a]+(\bar{n}+1)\gamma_{\rm eff}\mathcal{D}[\hat{\mathcal{S}}^{\dagger}_{\xi}\ah \hat{\mathcal{S}}_{\xi}](\rh_{a})\notag\\
    &+\bar{n}\gamma_{\rm eff}\mathcal{D}[\hat{\mathcal{S}}^{\dagger}_{\xi}\ad \hat{\mathcal{S}}_{\xi}](\rh_{a}),
\label{eqn:cavitymaster}
\end{align}
where $\rh_{a}$ is the density matrix of the system cavity. This master equation captures the dynamics of a cavity within a squeezed thermal reservoir, as visualized in the lower panel of Fig.~\ref{fig:BB}(a).

We focus on the regime where $\abs{\omega_{a,\rm eff}}>\abs{\Lambda}$, which ensures that the quadratic Hamiltonian remains stable and diagonalizable. In regimes where adiabatic elimination and Bogoliubov diagonalization no longer apply, one has to resort to other techniques such as bosonic third quantization~\cite{Prosen2008,McDonald2023}, which lies beyond the scope of the present work. Following the effective master equation~(\ref{eqn:cavitymaster}), the system reaches a steady state in the form of a squeezed thermal state, $\hat{\mathcal{S}}_{\xi_a}^\dag\rh_{\bar{n}_a}\hat{\mathcal{S}}_{\xi_a}$, characterized by a cavity squeezing parameter $\xi_{a}$ and a mean thermal occupation number $\bar{n}_a$, which can be tailored through the choice of microscopic parameters in the coupled cavity setup.  Details of the technical derivations of the cavity steady state can be found in Appendix~\ref{sec:rhoass}.

To examine the reservoir squeezing effect and its experimental relevance,  we simulate cavity dynamics under squeezed reservoir engineering using parameters motivated by coupled microwave cavities~\cite{Krantz2019, Miyanaga2021} while operating in the off-resonant and bad-cavity regime. Details of the numerical simulations can be found in Appendix~\ref{sec:cavitysim}. The simulated steady-state Wigner function of the system cavity~\cite{Cahill1969}, shown in Fig.~\ref{fig:BB}(b), exhibits a clear elliptical shape in phase space. This anisotropic distribution directly reflects the quadrature-dependent fluctuations induced by the engineered squeezed dissipation.  These numerical simulation results support the theoretical predictions of squeezed thermal reservoir engineering in the bosonic case.
The capability to stabilize steady-state squeezing in a cavity opens opportunities for enhanced quantum sensing and continuous-variable entanglement generation~\cite{Zippilli2015,Angeletti2023}.

\section{Conclusion}
We have introduced a general framework for engineering squeezed thermal reservoirs through passive linear coupling to a lossy auxiliary mode in a normal thermal environment. The central mechanism arises from interference between number-conserving and number-nonconserving interaction processes, giving rise to correlated dissipative dynamics characteristic of a squeezed reservoir. The generality of the framework is illustrated through two representative implementations: directional suppression of qubit decoherence and steady-state squeezing of a bosonic cavity mode, both achievable under experimentally accessible conditions.

Unlike conventional schemes that rely on driven nonlinear processes, externally injected squeezed fields, or time-dependent modulations, our approach realizes effective squeezed dissipation using only passive linear interactions widely available in existing quantum platforms, lowering the experimental overhead for squeezed reservoir engineering. Given the ubiquity of bilinear coupling in current quantum platforms, the mechanism developed here provides a broadly applicable and scalable route to engineered squeezed environments, with potential applications including dissipative squeezing, entanglement stabilization, and quantum heat engines.

The present analysis is formulated within the off-resonant or bad-cavity regime where adiabatic elimination leads to an effective Lindblad description. Extensions beyond this regime, including strong coupling and non-Markovian dynamics, remain open directions for future investigation.
This framework can also be extended to incorporate parametric effects, including those arising from optomechanical interactions or three-wave mixing, offering additional tunability for engineering open-system dynamics.
Future directions include exploring multimode generalizations and assessing the robustness of the engineered reservoir in noisy environments.
Together, these avenues broaden the scope of quantum control and support the development of scalable and resource-efficient strategies for quantum technologies.

\begin{acknowledgments}
We thank Shin-Tza Wu, Ping-Yi Wen, and Yen-Hsiang Lin for helpful discussions. This work was supported by the National Science and Technology Council, Taiwan, under Grants No.~111-2112-M-002-049-MY3, No.~111-2119-M-007-009, No.~112-2119-M-007-008, No.~113-2119-M-007-013, No.~114-2119-M-007-013, No.~114-2124-M-002-003, and No.~114-2112-M-002-021-MY3, and from the National Taiwan University, under Grants No.~111L74135, No.~112L7320, No.~112L7445, No.~114L895001, and No.~115L893701. C.-H.W. is also grateful for support from the Fubon Foundation, the Physics Division, National Center for Theoretical Sciences, Taiwan, and the Department of Physics, College of Science, National Taiwan University.
\end{acknowledgments}

\appendix

\section{Formal derivation of the effective master equation \label{AppendixA}}
We here derive the effective system dynamics using a standard master equation approach. Consider a general bilinear interaction between a quantum system $S$ and a dissipative bosonic mode $B$, with the total Hamiltonian given in Eq.~\eqref{eqn:Htot}. Coupling the bath mode to a thermal environment yields the Lindblad master equation for the full $S+B$ setup, given in Eq.~\eqref{eqn:totalmaster}.

In the off-resonant coupling regime, $\abs{ \omega_b - \omega_s} \gg g$, or in the bad-cavity limit, $\kappa \gg g$, we apply the method of adiabatic elimination~\cite{Gardiner2004,Jager2022,Jager2023}, assuming that the bath mode rapidly equilibrates to a thermal state.  This procedure is formally equivalent to a Schrieffer-Wolff transformation, with the generator $\mathcal{S} = \bd \hat{F} - \hat{F}^\dagger \bh$, where $\hat{F}$ is a system field operator that accounts for the interaction with the dissipative mode.

Following the treatment for eliminating lossy bosonic modes~\cite{Jager2022,Jager2023}, the system field operator takes the form
\begin{align} \hat{F} &= -\frac{g e^{i(\varphi_b - \varphi_s)}}{\omega_b - \omega_s - i\kappa/2} \sh - \frac{g e^{i(\varphi_b + \varphi_s)}}{\omega_b + \omega_s - i\kappa/2} \sd \notag\\&= e^{i (\varphi_b-\varphi_j)} \sqrt{\frac{\gamma_{\rm eff}}{\kappa}}\, \hat{J}_{\rm eff}. 
\label{eqn:F}
\end{align}
This field operator has the same structure as the jump operator $\hat{J}_{\rm eff}$ in Eq.~(\ref{eqn:Jeff}), up to a proportionality factor. Based on this relation, the effective master equation and Hamiltonian derived via adiabatic elimination can be compactly expressed in terms of $\hat{J}_{\rm eff}$, with the dissipative part matching the second-order interference picture established in Sec.~\ref{sec:sqpara}.

Specifically, the effective master equation reads
\begin{align}
    \dot{\rh}_s = &-\frac{i}{\hbar} [\hat{H}_{s, \rm eff}, \rh_s]+\left(\bar{n}+1\right)\gamma_{\rm eff}\mathcal{D}[\hat{J}_{\rm eff}]\left(\rh_s\right)\notag\\
    &+\bar{n}\gamma_{\rm eff}\mathcal{D}[\hat{J}_{\rm eff}^\dag]\left(\rh_s\right),
\end{align}
where the effective Hamiltonian is given by
\begin{align}
    \frac{1}{\hbar} \hat{H}_{s, \rm eff} =&\,\, \omega_s \sd \sh + \frac{g}{2} \sqrt{\frac{\gamma_{\rm eff}}{\kappa}}\left( \hat{J}_{\rm eff}^\dagger e^{i\varphi_j}(\sh e^{-i\varphi_s} \right. \notag\\&+ \left.\sd e^{i\varphi_s}) + (\sh e^{-i\varphi_s} + \sd e^{i\varphi_s}) \hat{J}_{\rm eff}e^{-i\varphi_j} \right) \notag\\
&+ \frac{g \bar{n}}{2}\sqrt{\frac{\gamma_{\rm eff}}{\kappa}} \left( [\hat{J}_{\rm eff}^\dagger e^{i\varphi_j}, \sh e^{-i\varphi_s} + \sd e^{i\varphi_s}] \right. \notag\\
& \left.+ [\sh e^{-i\varphi_s} + \sd e^{i\varphi_s}, \hat{J}_{\rm eff}e^{-i\varphi_j}] \right) + \bar{n} \omega_b.
\label{eqn:Heff}
\end{align}
This formal derivation yields the effective master equation and Hamiltonian used throughout the remainder of our analysis.

\section{Theory and simulation of the quantum Rabi model}
Here, we derive the effective qubit Hamiltonian under a squeezed thermal reservoir and present numerical simulations of the qubit dynamics, including additional results beyond the main text, to verify the predicted dependence of qubit dephasing on the system-bath detuning.

\subsection{Effective qubit Hamiltonian \label{sec:Hqeff}}
In the qubit implementation, we consider a two-level system (qubit) linearly coupled to a bosonic bath mode, described by the quantum Rabi model. Taking $\sh = \shm$ and $\varphi_s = \varphi_b = 0$, the effective Hamiltonian for the qubit is obtained by evaluating Eq.~(\ref{eqn:Heff}):
\begin{align}
\frac{1}{\hbar}\Hh_{q,\rm eff} 
=&\frac{\omega_q}{2} \shz + \frac{g}{2}\sqrt{\frac{\gamma_{\rm eff}}{\kappa}} \left(\Jh_{\rm eff}^\dag e^{i\varphi_j} + \Jh_{\rm eff} e^{-i\varphi_j}\right) \left(\shm + \shp\right) \notag\\
&+ \frac{g\bar{n}}{2} \sqrt{\frac{\gamma_{\rm eff}}{\kappa}} \left[\Jh_{\rm eff}^\dag e^{i\varphi_j} - \Jh_{\rm eff} e^{-i\varphi_j}, \shm + \shp\right] + \bar{n} \omega_b\notag\\
=&\, \left( \frac{\omega_q}{2} - \frac{\gamma_{\rm eff}(2\bar{n}+1)(\omega_b^2 - \omega_q^2 - \kappa^2/4)}{4\kappa\omega_b} \right) \shz \notag\\
=&\, \frac{\omega_{q,\rm eff}}{2} \shz,\\
\omega_{q,\rm eff}\equiv &\,\omega_q-\frac{\gamma_{\rm eff}(2\bar{n}+1)(\omega_b^2-\omega_q^2-\kappa^2 /4)}{2\kappa\omega_b},
\label{eqn:omegaqeff}
\end{align}
where constant terms have been omitted, as they do not influence the dynamics of the system. This effective Hamiltonian indicates a qubit frequency shift induced by the dissipative bath.

\subsection{Numerical simulation of qubit dynamics \label{sec:qubitsim}}
We simulate the qubit dynamics using both the full master equation~(\ref{eqn:totalmaster}) and the effective master equation~(\ref{eqn:qubitmaster}), implemented in Python using the mesolve function in the QuTiP package~\cite{Johansson2012,Johansson2013}. The qubit is initialized in states aligned along the $x$ and $y$ axes of the Bloch sphere, with $\hat{\rho}_0 = \frac{1}{2}(1 + \hat{\sigma}_x)$ and $\hat{\rho}_0 = \frac{1}{2}(1 + \hat{\sigma}_y)$, respectively, and the cavity mode is truncated at a photon number cutoff of 30. To assess the effect of detuning on squeezing, we vary the qubit frequency $\omega_q$ while fixing all other parameters as listed in Table~\ref{tab:QRM}. These parameters correspond to typical values in circuit QED experiments using fluxonium qubits operated in the dispersive regime~\cite{Manucharyan2009,Nguyen2019}.\\

\begin{center}
\begin{table}[htbp]
\caption{Simulation parameters and evaluated quantities for Figs.~\ref{fig:QRM} and~\ref{fig:QRMevo}. The qubit frequency $\omega_q$ is varied to explore the dependence of the engineered squeezing on the system-bath detuning, while all other parameters are kept fixed. The squeezing strength $r$, squeezing angle $\theta$, effective dissipation rate $\gamma$, and effective qubit frequency $\omega_{q,\rm eff}$ are calculated from the analytical expressions in Eqs.~\eqref{eqn:sqparam} and~\eqref{eqn:omegaqeff}.}
\begin{tabular}{cccc}
\\
\hline\hline
Figure(s) & \ref{fig:QRM}(b) and \ref{fig:QRMevo}(a) & \ref{fig:QRMevo}(b) & \ref{fig:QRMevo}(c) \\
\hline
$\omega_q/2\pi$ (MHz) & 100 & 200 & 300\\
$\omega_b/2\pi$ (MHz) & 1000 & 1000 & 1000\\
$\kappa/2\pi$ (MHz) & 600 & 600 & 600\\
$g/2\pi$ (MHz) & 176 & 176 & 176\\
$\bar{n}$ & 10 & 10 & 10\\
$\omega_{q,\rm eff}/2\pi$ (MHz) & $-0.08$ & $-2.68$ & $-10.00$\\
$\gamma_{\rm eff}/2\pi$ (MHz) & 6.35 & 13.31 & 21.60\\
$r$ & 1.19 & 0.85 & 0.65\\
$\theta$ (deg) & 176.82 & 173.48 & 169.80\\
\hline\hline
\end{tabular}
\label{tab:QRM}
\end{table}
\end{center}

\begin{figure}[htbp]
\begin{center}
\includegraphics[width=\linewidth]{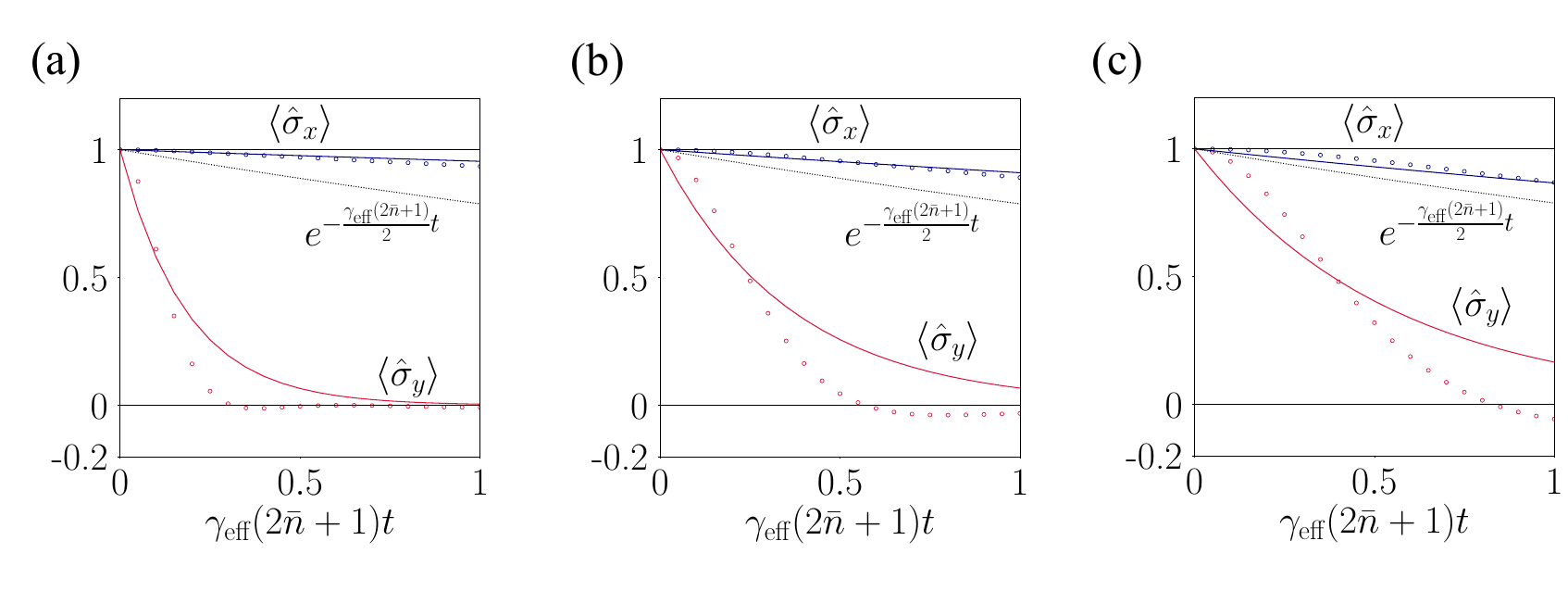}
\caption{
Squeezed qubit dynamics under varying detuning.  
Simulated evolution of $\langle \hat{\sigma}_x \rangle$ and $\langle \hat{\sigma}_y \rangle$, starting from initial states aligned along the $x$ and $y$ axes, under engineered squeezed dissipation for three different qubit frequencies: (a) $\omega_q / 2 \pi = 100$~MHz, (b) $\omega_q / 2 \pi = 200$~MHz, and (c) $\omega_q / 2 \pi = 300$~MHz.  
Solid lines represent the effective master equation~(\ref{eqn:qubitmaster}), and circles indicate results from the full model~(\ref{eqn:totalmaster}); the dotted line shows decay under a normal thermal reservoir for comparison.
Simulation parameters are summarized in Table~\ref{tab:QRM}. The simulations reveal suppression of qubit decoherence along the $x$ axis (squeezed direction), with decreasing quadrature anisotropy at smaller detuning.
}
\label{fig:QRMevo} 
\end{center}
\end{figure}

Figure~\ref{fig:QRMevo} shows the simulated qubit dynamics under varying detuning. This chosen parameter configuration realizes a strongly squeezed reservoir in the far-detuned regime ($\omega_q \ll \omega_b$) of panel (a), where the effective squeezing strength is the highest among the three panels. As the qubit frequency increases in panels (b) and (c), the system moves closer to resonance with the bath mode, leading to a gradual reduction in squeezing strength. These observations are consistent with the analytical expression for the squeezing parameter [Eq.~(\ref{eqn:sqparam})] and the numerical results plotted in Fig.~\ref{fig:sqparameter}.

\section{Theory and simulation of the bosonic interaction model}
Here, we derive the effective Hamiltonian for the system cavity under linear coupling to a dissipative mode, obtain its steady state as a squeezed thermal state, and present numerical simulations, including additional results beyond the main text, to verify the predicted steady-state squeezing phenomena.

\subsection{Effective cavity Hamiltonian \label{sec:Haeff}}
In the cavity implementation, we consider two linearly coupled bosonic (cavity) modes, where one mode acts as the system mode and the other as a dissipative bath mode, described by the bosonic interaction model. Taking $\sh = \ah$ and $\varphi_a = \varphi_b = 0$, the effective Hamiltonian of the system cavity is obtained by evaluating Eq.~(\ref{eqn:Heff}):
\begin{align}
\frac{1}{\hbar}\Hh_{a,\rm eff}=&\left(\omega_a-\frac{\gamma_{\rm eff}(\omega_b^2-\omega_a^2+\kappa^2 /4)}{2\kappa\omega_a}\right)\ad\ah\notag\\
&-\frac{g^2}{2}\left(\frac{1}{\omega_b-\omega_a+i\kappa /2}+\frac{1}{\omega_b+\omega_a-i\kappa /2}\right)\hat{a}^{\dag 2}\notag\\
&-\frac{g^2}{2}\left(\frac{1}{\omega_b-\omega_a-i\kappa /2}+\frac{1}{\omega_b+\omega_a+i\kappa /2}\right)\ah^2\notag\\
\quad=&\, \omega_{a,\rm eff} \ad \ah + \frac{i}{2} (\Lambda \hat{a}^{\dagger 2} -\Lambda^{*} \ah^2),\notag\\
\omega_{a,\rm eff}\equiv& \, \omega_a-\frac{\gamma_{\rm eff}(\omega_b^2-\omega_a^2+\kappa^2 /4)}{2\kappa\omega_a}, \,\notag\\
\Lambda\equiv& \, ig^2\left(\frac{1}{\omega_b-\omega_a+i\kappa /2}+\frac{1}{\omega_b+\omega_a-i\kappa /2}\right),
\label{eqn:omegalambda}
\end{align}
where constant terms have been omitted, as they do not influence the dynamics of the system. Compared to the quantum Rabi model, this effective Hamiltonian is independent of the mean thermal occupation number $\bar{n}$. Moreover, the coherent two-photon terms  $\ah^2$ and $\hat{a}^{\dag 2}$ generate additional unitary squeezing, supplementing the dissipative squeezing established through engineered loss.

\subsection{Characterization of the cavity steady state \label{sec:rhoass}}
In the regime where $\abs{\omega_{a,\rm{eff}}} >\abs{\Lambda}$, the effective Hamiltonian may be diagonalized by the squeezing operator $\hat{\mathcal{S}}_{\xi_d}=\exp\left[\frac{1}{2}\left(\xi_d^{*}\ah^2-\xi_d\hat{a}^{\dag 2}\right)\right]$, where $\hat{\mathcal{S}}_{\xi_d}^\dag\ah\hat{\mathcal{S}}_{\xi_d}=\ah\cosh(r_d)-\ad e^{i\theta_d}\sinh(r_d)$. Applying this Bogoliubov diagonalization, the effective master equation~(\ref{eqn:cavitymaster}) takes the form
\begin{align}
\dot{\rh}_a=&-i\left[\Omega\hat{\mathcal{S}}_{\xi_d}\ad\ah\hat{\mathcal{S}}_{\xi_d}^\dag,\rh_a\right]+\left(\bar{n}+1\right)\gamma_{\rm eff}\mathcal{D}\left[\hat{\mathcal{S}}_\xi^\dag\ah\hat{\mathcal{S}}_\xi\right]\left(\rh_a\right)\notag\\
&+\bar{n}\gamma_{\rm eff}\mathcal{D}\left[\hat{\mathcal{S}}_\xi^\dag\ad\hat{\mathcal{S}}_\xi\right]\left(\rh_a\right).
\label{eqn:cavitymasterd}
\end{align}\noindent
Note that $r_d=\frac{1}{2}\tanh^{-1}\left(\frac{\abs{\Lambda}}{\omega_{a,\rm eff}}\right)$, $\theta_d=\arg\left(i\Lambda\right)=\pi+\arg\left(\frac{\omega_b-\omega_a-i\kappa/2}{\omega_b+\omega_a-i\kappa/2}\right)$, and $\Omega=\sqrt{\omega_{a, \rm eff}^2-\abs{\Lambda}^2}$.

To find the steady state of Eq.~(\ref{eqn:cavitymasterd}), we first apply a squeezing transformation to the system density operator,  $\rt_a=\hat{\mathcal{S}}_\xi\rh_a\hat{\mathcal{S}}_\xi^\dag$, such that the environmental squeezing is transferred into the transformed Hamiltonian, and the dissipator becomes thermal. The transformed master equation becomes
\begin{align}
\dot{\rt}_a=&-i\left[\Omega\hat{\mathcal{S}}_{\xi+\xi_d}\ad\ah\hat{\mathcal{S}}_{\xi+\xi_d}^\dag,\rt_a\right]+\left(\bar{n}+1\right)\gamma_{\rm eff}\mathcal{D}\left[\ah\right]\left(\rt_a\right)
\notag\\
&+\bar{n}\gamma_{\rm eff}\mathcal{D}\left[\ad\right]\left(\rt_a\right),
\label{eqn:cavitymastert}
\end{align}\noindent
Note that $\xi$ and $\xi_d$ share the same angle ($\theta = \theta_d$), the corresponding squeezing operators commute, and the parameters can be added directly.

The steady-state solution of Eq.~\eqref{eqn:cavitymastert} in the transformed frame takes the form of a squeezed thermal state, $\rt_{a,\rm ss}=\hat{\mathcal{S}}_{\xi_t}\rh_{\bar{n}_a}\hat{\mathcal{S}}_{\xi_t}^\dag$, with
\begin{align}
r_t&=\frac{1}{2}\tanh^{-1}\left(\sqrt{\frac{\sinh^2[2(r+r_d)]}{\cosh^2[2(r+r_d)]+\frac{\gamma_{\rm eff}^2}{4\Omega^2}}}\right),\notag\\ 
\theta_t&=\theta+\tan^{-1}\left(\frac{\gamma_{\rm eff}}{2\Omega\cosh[2(r+r_d)]}\right),
\end{align}\noindent
and $\rh_{\bar{n}_a}=\frac{1}{\bar{n}_a+1}\left(\frac{\bar{n}_a}{\bar{n}_a+1}\right)^{\ad\ah}$ represents a thermal state with a mean thermal occupation number $\bar{n}_a$ given by
\begin{align}
\bar{n}_a=\frac{1}{2}\left[\left(2\bar{n}+1\right)\cosh(2r_t)-1\right].
\label{eqn:na}
\end{align}

Applying the inverse squeezing transformation $\hat{\mathcal{S}}_\xi^\dag$ returns the system to the original frame, where the steady state of the system cavity becomes
\begin{align}
\rh_{a,\rm{ss}} = \frac{1}{\bar{n}_a + 1} 
\hat{\mathcal{S}}_\xi^\dag \hat{\mathcal{S}}_{\xi_t} 
\left( \frac{\bar{n}_a}{\bar{n}_a + 1} \right)^{\ad\ah}
\hat{\mathcal{S}}_{\xi_t}^\dag \hat{\mathcal{S}}_\xi.
\label{eqn:thoass}
\end{align} Combining the effects of both squeezing operations, we obtain the steady state of Eq.~\eqref{eqn:cavitymaster} as a squeezed thermal state, 
\begin{align}
\rh_{a,\rm ss}=\hat{\mathcal{S}}_{\xi_{a}}^\dag\rh_{\bar{n}_a}\hat{\mathcal{S}}_{\xi_{a}}, \, \xi_{a}=r_a e^{i\theta_a},
\end{align}
where
\begin{align}
r_a=&\tanh^{-1}\abs{\frac{\tanh(r)-e^{i\left(\theta_t-\theta\right)}\tanh(r_t)}{1-e^{i\left(\theta_t-\theta\right)}\tanh(r)\tanh(r_t)}},\notag\\
\theta_a=&\pi+\theta\notag\\
&\phantom{\pi}+\tan^{-1}\left(\tfrac{\sin(\theta-\theta_t)\sinh(2r_t)}{\sinh(2r)\cosh(2r_t)-\cos(\theta-\theta_t)\cosh(2r)\sinh(2r_t)}\right).
\label{eqn:xia}
\end{align}
These expressions fully characterize the steady-state squeezing properties of the system cavity, controlled by the microscopic parameters of the coupled cavity-cavity setup.

\subsection{Numerical simulation of cavity steady state \label{sec:cavitysim}}
We simulate the cavity dynamics under an engineered squeezed reservoir by solving for the steady state of the effective master equation~(\ref{eqn:cavitymaster}) using the steady state function in the QuTiP package~\cite{Johansson2012,Johansson2013}. To examine the dependence of steady-state squeezing on detuning, we vary the system cavity frequency as listed in Table~\ref{tab:ss}, while keeping all other parameters fixed. The chosen parameters correspond to the strong coupling and bad-cavity regime between superconducting microwave resonators~\cite{Krantz2019, Miyanaga2021}, and the cavity mode is truncated at a photon number cutoff of 200.

\begin{center}
\begin{table}[htbp]
\caption{Simulation parameters and evaluated quantities for Figs.~\ref{fig:BB} and~\ref{fig:ss}. The system cavity frequency $\omega_a$ is varied to examine how the steady-state squeezing depends on detuning, while all other parameters are kept fixed. The effective cavity frequency $\omega_{a,\rm eff}$ and Hamiltonian squeezing strength $\Lambda$ are calculated from Eq.~\eqref{eqn:omegalambda}. The resulting squeezing strength $r_a$, squeezing angle $\theta_a$, and mean thermal occupation number $\bar{n}_a$ of the system cavity steady state are calculated from Eqs.~(\ref{eqn:na}) and~(\ref{eqn:xia}).}
\begin{tabular}{cccc}
\\
\hline\hline
Figure(s) & \ref{fig:BB}(b) and \ref{fig:ss}(a) & \ref{fig:ss}(b) & \ref{fig:ss}(c)\\
\hline
$\omega_a / 2\pi$ (GHz) & 1.12 & 1.20 & 1.30 \\
$\omega_b / 2\pi$ (GHz) & 4.4 & 4.4 & 4.4 \\
$g / 2\pi$ (GHz) & 1.2 & 1.2 & 1.2 \\
$\kappa / 2 \pi$ (GHz) & 4.0 & 4.0 & 4.0 \\
$\bar{n}$ & 0 & 0 & 0 \\
$\omega_{a,\rm eff} / 2\pi$ (GHz) & 0.57 & 0.65 & 0.75 \\
$\Lambda / 2\pi$ (GHz) & $0.11 + 0.55i$ & $0.12 + 0.55i$ & $0.13 + 0.55i$ \\
$r_a$ & 1.03 & 0.64 & 0.49 \\
$\theta_a$ (deg) & 359.98 & 359.88 & 359.72 \\
$\bar{n}_a$ & 8.85 & 3.50 & 2.23 \\
\hline\hline
\end{tabular}
\label{tab:ss}
\end{table}
\end{center}

\begin{figure}[htbp]
\centering
\includegraphics[width=\linewidth]{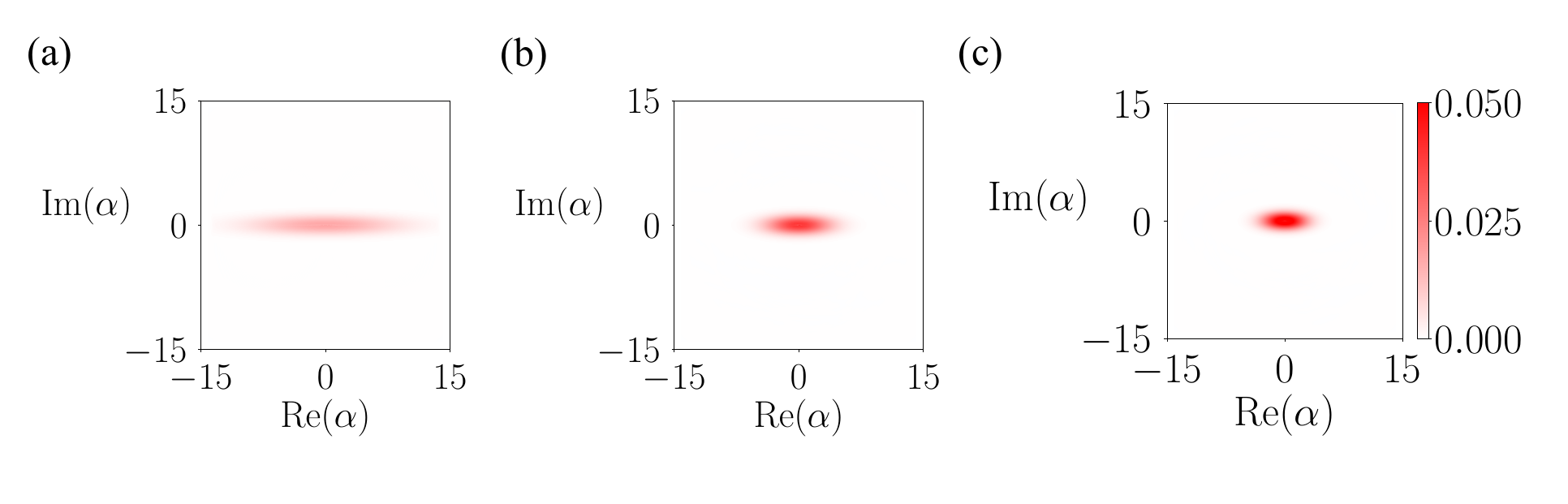}
\caption{Steady-state Wigner functions of the system cavity under engineered squeezed dissipation. Simulated Wigner functions are shown for three different cavity frequencies: (a) $\omega_a / 2 \pi = 1.12$~GHz, (b) $\omega_a / 2 \pi = 1.20$~GHz, and (c) $\omega_a / 2 \pi = 1.30$~GHz. All panels share a common color normalization for direct visual comparison. Simulated parameters are summarized in Table~\ref{tab:ss}. The elliptical shapes in phase space reflect quadrature-dependent fluctuations associated with engineered squeezed dissipation.}
\label{fig:ss}
\end{figure}

Figure~\ref{fig:ss} shows the simulated steady-state Wigner functions of the system cavity for three different values of $\omega_a$. In panel~(a), the reservoir is strongly squeezed, resulting in a pronounced elliptical distribution in phase space. As detuning decreases in panels~(b) and (c), the squeezing strength decreases, and the Wigner function becomes increasingly circular. These simulations confirm the controllability of steady-state squeezing via system parameters in our reservoir engineering framework.

\bibliography{SqueezeRef}

\begin{thebibliography}{75}%
\makeatletter
\providecommand \@ifxundefined [1]{%
 \@ifx{#1\undefined}
}%
\providecommand \@ifnum [1]{%
 \ifnum #1\expandafter \@firstoftwo
 \else \expandafter \@secondoftwo
 \fi
}%
\providecommand \@ifx [1]{%
 \ifx #1\expandafter \@firstoftwo
 \else \expandafter \@secondoftwo
 \fi
}%
\providecommand \natexlab [1]{#1}%
\providecommand \enquote  [1]{``#1''}%
\providecommand \bibnamefont  [1]{#1}%
\providecommand \bibfnamefont [1]{#1}%
\providecommand \citenamefont [1]{#1}%
\providecommand \href@noop [0]{\@secondoftwo}%
\providecommand \href [0]{\begingroup \@sanitize@url \@href}%
\providecommand \@href[1]{\@@startlink{#1}\@@href}%
\providecommand \@@href[1]{\endgroup#1\@@endlink}%
\providecommand \@sanitize@url [0]{\catcode `\\12\catcode `\$12\catcode
  `\&12\catcode `\#12\catcode `\^12\catcode `\_12\catcode `\%12\relax}%
\providecommand \@@startlink[1]{}%
\providecommand \@@endlink[0]{}%
\providecommand \url  [0]{\begingroup\@sanitize@url \@url }%
\providecommand \@url [1]{\endgroup\@href {#1}{\urlprefix }}%
\providecommand \urlprefix  [0]{URL }%
\providecommand \Eprint [0]{\href }%
\providecommand \doibase [0]{https://doi.org/}%
\providecommand \selectlanguage [0]{\@gobble}%
\providecommand \bibinfo  [0]{\@secondoftwo}%
\providecommand \bibfield  [0]{\@secondoftwo}%
\providecommand \translation [1]{[#1]}%
\providecommand \BibitemOpen [0]{}%
\providecommand \bibitemStop [0]{}%
\providecommand \bibitemNoStop [0]{.\EOS\space}%
\providecommand \EOS [0]{\spacefactor3000\relax}%
\providecommand \BibitemShut  [1]{\csname bibitem#1\endcsname}%
\let\auto@bib@innerbib\@empty
\bibitem [{\citenamefont {Poyatos}\ \emph {et~al.}(1996)\citenamefont
  {Poyatos}, \citenamefont {Cirac},\ and\ \citenamefont
  {Zoller}}]{Poyatos1996}%
  \BibitemOpen
  \bibfield  {author} {\bibinfo {author} {\bibfnamefont {J.~F.}\ \bibnamefont
  {Poyatos}}, \bibinfo {author} {\bibfnamefont {J.~I.}\ \bibnamefont {Cirac}},\
  and\ \bibinfo {author} {\bibfnamefont {P.}~\bibnamefont {Zoller}},\
  }\bibfield  {title} {\bibinfo {title} {{Quantum reservoir engineering with
  laser cooled trapped ions}},\ }\href
  {https://doi.org/10.1103/PhysRevLett.77.4728} {\bibfield  {journal} {\bibinfo
   {journal} {Phys. Rev. Lett.}\ }\textbf {\bibinfo {volume} {77}},\ \bibinfo
  {pages} {4728} (\bibinfo {year} {1996})}\BibitemShut {NoStop}%
\bibitem [{\citenamefont {Verstraete}\ \emph {et~al.}(2009)\citenamefont
  {Verstraete}, \citenamefont {Wolf},\ and\ \citenamefont
  {Cirac}}]{Verstraete2009}%
  \BibitemOpen
  \bibfield  {author} {\bibinfo {author} {\bibfnamefont {F.}~\bibnamefont
  {Verstraete}}, \bibinfo {author} {\bibfnamefont {M.~M.}\ \bibnamefont
  {Wolf}},\ and\ \bibinfo {author} {\bibfnamefont {J.~I.}\ \bibnamefont
  {Cirac}},\ }\bibfield  {title} {\bibinfo {title} {{Quantum computation and
  quantum-state engineering driven by dissipation}},\ }\href
  {https://doi.org/10.1038/nphys1342} {\bibfield  {journal} {\bibinfo
  {journal} {Nat. Phys.}\ }\textbf {\bibinfo {volume} {5}},\ \bibinfo {pages}
  {633} (\bibinfo {year} {2009})}\BibitemShut {NoStop}%
\bibitem [{\citenamefont {Carvalho}\ \emph {et~al.}(2001)\citenamefont
  {Carvalho}, \citenamefont {Milman}, \citenamefont {{De Matos Filho}},\ and\
  \citenamefont {Davidovich}}]{Carvalho2001}%
  \BibitemOpen
  \bibfield  {author} {\bibinfo {author} {\bibfnamefont {A.~R.~R.}\
  \bibnamefont {Carvalho}}, \bibinfo {author} {\bibfnamefont {P.}~\bibnamefont
  {Milman}}, \bibinfo {author} {\bibfnamefont {R.~L.}\ \bibnamefont {{De Matos
  Filho}}},\ and\ \bibinfo {author} {\bibfnamefont {L.}~\bibnamefont
  {Davidovich}},\ }\bibfield  {title} {\bibinfo {title} {{Decoherence, pointer
  engineering, and quantum state protection}},\ }\href
  {https://doi.org/10.1103/PhysRevLett.86.4988} {\bibfield  {journal} {\bibinfo
   {journal} {Phys. Rev. Lett.}\ }\textbf {\bibinfo {volume} {86}},\ \bibinfo
  {pages} {4988} (\bibinfo {year} {2001})}\BibitemShut {NoStop}%
\bibitem [{\citenamefont {Krauter}\ \emph {et~al.}(2011)\citenamefont
  {Krauter}, \citenamefont {Muschik}, \citenamefont {Jensen}, \citenamefont
  {Wasilewski}, \citenamefont {Petersen}, \citenamefont {Cirac},\ and\
  \citenamefont {Polzik}}]{Krauter2011}%
  \BibitemOpen
  \bibfield  {author} {\bibinfo {author} {\bibfnamefont {H.}~\bibnamefont
  {Krauter}}, \bibinfo {author} {\bibfnamefont {C.~A.}\ \bibnamefont
  {Muschik}}, \bibinfo {author} {\bibfnamefont {K.}~\bibnamefont {Jensen}},
  \bibinfo {author} {\bibfnamefont {W.}~\bibnamefont {Wasilewski}}, \bibinfo
  {author} {\bibfnamefont {J.~M.}\ \bibnamefont {Petersen}}, \bibinfo {author}
  {\bibfnamefont {J.~I.}\ \bibnamefont {Cirac}},\ and\ \bibinfo {author}
  {\bibfnamefont {E.~S.}\ \bibnamefont {Polzik}},\ }\bibfield  {title}
  {\bibinfo {title} {{Entanglement generated by dissipation and steady state
  entanglement of two macroscopic objects}},\ }\href
  {https://doi.org/10.1103/PhysRevLett.107.080503} {\bibfield  {journal}
  {\bibinfo  {journal} {Phys. Rev. Lett.}\ }\textbf {\bibinfo {volume} {107}},\
  \bibinfo {pages} {080503} (\bibinfo {year} {2011})}\BibitemShut {NoStop}%
\bibitem [{\citenamefont {Lin}\ \emph {et~al.}(2013)\citenamefont {Lin},
  \citenamefont {Gaebler}, \citenamefont {Reiter}, \citenamefont {Tan},
  \citenamefont {Bowler}, \citenamefont {S{\o}rensen}, \citenamefont
  {Leibfried},\ and\ \citenamefont {Wineland}}]{Lin2013}%
  \BibitemOpen
  \bibfield  {author} {\bibinfo {author} {\bibfnamefont {Y.}~\bibnamefont
  {Lin}}, \bibinfo {author} {\bibfnamefont {J.~P.}\ \bibnamefont {Gaebler}},
  \bibinfo {author} {\bibfnamefont {F.}~\bibnamefont {Reiter}}, \bibinfo
  {author} {\bibfnamefont {T.~R.}\ \bibnamefont {Tan}}, \bibinfo {author}
  {\bibfnamefont {R.}~\bibnamefont {Bowler}}, \bibinfo {author} {\bibfnamefont
  {A.~S.}\ \bibnamefont {S{\o}rensen}}, \bibinfo {author} {\bibfnamefont
  {D.}~\bibnamefont {Leibfried}},\ and\ \bibinfo {author} {\bibfnamefont
  {D.~J.}\ \bibnamefont {Wineland}},\ }\bibfield  {title} {\bibinfo {title}
  {{Dissipative production of a maximally entangled steady state of two quantum
  bits}},\ }\href {https://doi.org/10.1038/nature12801} {\bibfield  {journal}
  {\bibinfo  {journal} {Nature (London)}\ }\textbf {\bibinfo {volume} {504}},\
  \bibinfo {pages} {415} (\bibinfo {year} {2013})}\BibitemShut {NoStop}%
\bibitem [{\citenamefont {Shankar}\ \emph {et~al.}(2013)\citenamefont
  {Shankar}, \citenamefont {Hatridge}, \citenamefont {Leghtas}, \citenamefont
  {Sliwa}, \citenamefont {Narla}, \citenamefont {Vool}, \citenamefont {Girvin},
  \citenamefont {Frunzio}, \citenamefont {Mirrahimi},\ and\ \citenamefont
  {Devoret}}]{Shankar2013}%
  \BibitemOpen
  \bibfield  {author} {\bibinfo {author} {\bibfnamefont {S.}~\bibnamefont
  {Shankar}}, \bibinfo {author} {\bibfnamefont {M.}~\bibnamefont {Hatridge}},
  \bibinfo {author} {\bibfnamefont {Z.}~\bibnamefont {Leghtas}}, \bibinfo
  {author} {\bibfnamefont {K.~M.}\ \bibnamefont {Sliwa}}, \bibinfo {author}
  {\bibfnamefont {A.}~\bibnamefont {Narla}}, \bibinfo {author} {\bibfnamefont
  {U.}~\bibnamefont {Vool}}, \bibinfo {author} {\bibfnamefont {S.~M.}\
  \bibnamefont {Girvin}}, \bibinfo {author} {\bibfnamefont {L.}~\bibnamefont
  {Frunzio}}, \bibinfo {author} {\bibfnamefont {M.}~\bibnamefont {Mirrahimi}},\
  and\ \bibinfo {author} {\bibfnamefont {M.~H.}\ \bibnamefont {Devoret}},\
  }\bibfield  {title} {\bibinfo {title} {{Autonomously stabilized entanglement
  between two superconducting quantum bits}},\ }\href
  {https://doi.org/10.1038/nature12802} {\bibfield  {journal} {\bibinfo
  {journal} {Nature (London)}\ }\textbf {\bibinfo {volume} {504}},\ \bibinfo
  {pages} {419} (\bibinfo {year} {2013})}\BibitemShut {NoStop}%
\bibitem [{\citenamefont {Pastawski}\ \emph {et~al.}(2011)\citenamefont
  {Pastawski}, \citenamefont {Clemente},\ and\ \citenamefont
  {Cirac}}]{Pastawski2011}%
  \BibitemOpen
  \bibfield  {author} {\bibinfo {author} {\bibfnamefont {F.}~\bibnamefont
  {Pastawski}}, \bibinfo {author} {\bibfnamefont {L.}~\bibnamefont
  {Clemente}},\ and\ \bibinfo {author} {\bibfnamefont {J.~I.}\ \bibnamefont
  {Cirac}},\ }\bibfield  {title} {\bibinfo {title} {{Quantum memories based on
  engineered dissipation}},\ }\href
  {https://doi.org/10.1103/PhysRevA.83.012304} {\bibfield  {journal} {\bibinfo
  {journal} {Phys. Rev. A}\ }\textbf {\bibinfo {volume} {83}},\ \bibinfo
  {pages} {012304} (\bibinfo {year} {2011})}\BibitemShut {NoStop}%
\bibitem [{\citenamefont {Goldstein}\ \emph {et~al.}(2011)\citenamefont
  {Goldstein}, \citenamefont {Cappellaro}, \citenamefont {Maze}, \citenamefont
  {Hodges}, \citenamefont {Jiang}, \citenamefont {S{\o}rensen},\ and\
  \citenamefont {Lukin}}]{Goldstein2011}%
  \BibitemOpen
  \bibfield  {author} {\bibinfo {author} {\bibfnamefont {G.}~\bibnamefont
  {Goldstein}}, \bibinfo {author} {\bibfnamefont {P.}~\bibnamefont
  {Cappellaro}}, \bibinfo {author} {\bibfnamefont {J.~R.}\ \bibnamefont
  {Maze}}, \bibinfo {author} {\bibfnamefont {J.~S.}\ \bibnamefont {Hodges}},
  \bibinfo {author} {\bibfnamefont {L.}~\bibnamefont {Jiang}}, \bibinfo
  {author} {\bibfnamefont {A.~S.}\ \bibnamefont {S{\o}rensen}},\ and\ \bibinfo
  {author} {\bibfnamefont {M.~D.}\ \bibnamefont {Lukin}},\ }\bibfield  {title}
  {\bibinfo {title} {{Environment-assisted precision measurement}},\ }\href
  {https://doi.org/10.1103/PhysRevLett.106.140502} {\bibfield  {journal}
  {\bibinfo  {journal} {Phys. Rev. Lett.}\ }\textbf {\bibinfo {volume} {106}},\
  \bibinfo {pages} {140502} (\bibinfo {year} {2011})}\BibitemShut {NoStop}%
\bibitem [{\citenamefont {Reiter}\ \emph {et~al.}(2017)\citenamefont {Reiter},
  \citenamefont {S{\o}rensen}, \citenamefont {Zoller},\ and\ \citenamefont
  {Muschik}}]{Reiter2017}%
  \BibitemOpen
  \bibfield  {author} {\bibinfo {author} {\bibfnamefont {F.}~\bibnamefont
  {Reiter}}, \bibinfo {author} {\bibfnamefont {A.~S.}\ \bibnamefont
  {S{\o}rensen}}, \bibinfo {author} {\bibfnamefont {P.}~\bibnamefont
  {Zoller}},\ and\ \bibinfo {author} {\bibfnamefont {C.~A.}\ \bibnamefont
  {Muschik}},\ }\bibfield  {title} {\bibinfo {title} {{Dissipative quantum
  error correction and application to quantum sensing with trapped ions}},\
  }\href {https://doi.org/10.1038/s41467-017-01895-5} {\bibfield  {journal}
  {\bibinfo  {journal} {Nat. Commun.}\ }\textbf {\bibinfo {volume} {8}},\
  \bibinfo {pages} {1822} (\bibinfo {year} {2017})}\BibitemShut {NoStop}%
\bibitem [{\citenamefont {Kapit}\ \emph {et~al.}(2014)\citenamefont {Kapit},
  \citenamefont {Hafezi},\ and\ \citenamefont {Simon}}]{Kapit2014}%
  \BibitemOpen
  \bibfield  {author} {\bibinfo {author} {\bibfnamefont {E.}~\bibnamefont
  {Kapit}}, \bibinfo {author} {\bibfnamefont {M.}~\bibnamefont {Hafezi}},\ and\
  \bibinfo {author} {\bibfnamefont {S.~H.}\ \bibnamefont {Simon}},\ }\bibfield
  {title} {\bibinfo {title} {{Induced self-stabilization in fractional quantum
  Hall states of light}},\ }\href {https://doi.org/10.1103/PhysRevX.4.031039}
  {\bibfield  {journal} {\bibinfo  {journal} {Phys. Rev. X}\ }\textbf {\bibinfo
  {volume} {4}},\ \bibinfo {pages} {031039} (\bibinfo {year}
  {2014})}\BibitemShut {NoStop}%
\bibitem [{\citenamefont {Wang}\ \emph {et~al.}(2019)\citenamefont {Wang},
  \citenamefont {Gullans}, \citenamefont {Porto}, \citenamefont {Phillips},\
  and\ \citenamefont {Taylor}}]{Wang2019}%
  \BibitemOpen
  \bibfield  {author} {\bibinfo {author} {\bibfnamefont {C.~H.}\ \bibnamefont
  {Wang}}, \bibinfo {author} {\bibfnamefont {M.~J.}\ \bibnamefont {Gullans}},
  \bibinfo {author} {\bibfnamefont {J.~V.}\ \bibnamefont {Porto}}, \bibinfo
  {author} {\bibfnamefont {W.~D.}\ \bibnamefont {Phillips}},\ and\ \bibinfo
  {author} {\bibfnamefont {J.~M.}\ \bibnamefont {Taylor}},\ }\bibfield  {title}
  {\bibinfo {title} {{Theory of Bose condensation of light via laser cooling of
  atoms}},\ }\href {https://doi.org/10.1103/PhysRevA.99.031801} {\bibfield
  {journal} {\bibinfo  {journal} {Phys. Rev. A}\ }\textbf {\bibinfo {volume}
  {99}},\ \bibinfo {pages} {031801} (\bibinfo {year} {2019})}\BibitemShut
  {NoStop}%
\bibitem [{\citenamefont {Puri}\ \emph {et~al.}(2019)\citenamefont {Puri},
  \citenamefont {Grimm}, \citenamefont {Campagne-Ibarcq}, \citenamefont
  {Eickbusch}, \citenamefont {Noh}, \citenamefont {Roberts}, \citenamefont
  {Jiang}, \citenamefont {Mirrahimi}, \citenamefont {Devoret},\ and\
  \citenamefont {Girvin}}]{Puri2019}%
  \BibitemOpen
  \bibfield  {author} {\bibinfo {author} {\bibfnamefont {S.}~\bibnamefont
  {Puri}}, \bibinfo {author} {\bibfnamefont {A.}~\bibnamefont {Grimm}},
  \bibinfo {author} {\bibfnamefont {P.}~\bibnamefont {Campagne-Ibarcq}},
  \bibinfo {author} {\bibfnamefont {A.}~\bibnamefont {Eickbusch}}, \bibinfo
  {author} {\bibfnamefont {K.}~\bibnamefont {Noh}}, \bibinfo {author}
  {\bibfnamefont {G.}~\bibnamefont {Roberts}}, \bibinfo {author} {\bibfnamefont
  {L.}~\bibnamefont {Jiang}}, \bibinfo {author} {\bibfnamefont
  {M.}~\bibnamefont {Mirrahimi}}, \bibinfo {author} {\bibfnamefont {M.~H.}\
  \bibnamefont {Devoret}},\ and\ \bibinfo {author} {\bibfnamefont {S.~M.}\
  \bibnamefont {Girvin}},\ }\bibfield  {title} {\bibinfo {title} {{Stabilized
  cat in a driven nonlinear cavity: A fault-tolerant error syndrome
  detector}},\ }\href {https://doi.org/10.1103/PhysRevX.9.041009} {\bibfield
  {journal} {\bibinfo  {journal} {Phys. Rev. X}\ }\textbf {\bibinfo {volume}
  {9}},\ \bibinfo {pages} {041009} (\bibinfo {year} {2019})}\BibitemShut
  {NoStop}%
\bibitem [{\citenamefont {Gertler}\ \emph {et~al.}(2021)\citenamefont
  {Gertler}, \citenamefont {Baker}, \citenamefont {Li}, \citenamefont {Shirol},
  \citenamefont {Koch},\ and\ \citenamefont {Wang}}]{Gertler2021}%
  \BibitemOpen
  \bibfield  {author} {\bibinfo {author} {\bibfnamefont {J.~M.}\ \bibnamefont
  {Gertler}}, \bibinfo {author} {\bibfnamefont {B.}~\bibnamefont {Baker}},
  \bibinfo {author} {\bibfnamefont {J.}~\bibnamefont {Li}}, \bibinfo {author}
  {\bibfnamefont {S.}~\bibnamefont {Shirol}}, \bibinfo {author} {\bibfnamefont
  {J.}~\bibnamefont {Koch}},\ and\ \bibinfo {author} {\bibfnamefont
  {C.}~\bibnamefont {Wang}},\ }\bibfield  {title} {\bibinfo {title}
  {{Protecting a bosonic qubit with autonomous quantum error correction}},\
  }\href {https://doi.org/10.1038/s41586-021-03257-0} {\bibfield  {journal}
  {\bibinfo  {journal} {Nature (London)}\ }\textbf {\bibinfo {volume} {590}},\
  \bibinfo {pages} {243} (\bibinfo {year} {2021})}\BibitemShut {NoStop}%
\bibitem [{\citenamefont {Barra}(2019)}]{Barra2019}%
  \BibitemOpen
  \bibfield  {author} {\bibinfo {author} {\bibfnamefont {F.}~\bibnamefont
  {Barra}},\ }\bibfield  {title} {\bibinfo {title} {{Dissipative charging of a
  quantum battery}},\ }\href {https://doi.org/10.1103/PhysRevLett.122.210601}
  {\bibfield  {journal} {\bibinfo  {journal} {Phys. Rev. Lett.}\ }\textbf
  {\bibinfo {volume} {122}},\ \bibinfo {pages} {210601} (\bibinfo {year}
  {2019})}\BibitemShut {NoStop}%
\bibitem [{\citenamefont {Wang}\ and\ \citenamefont
  {Gertler}(2019)}]{Wang2019a}%
  \BibitemOpen
  \bibfield  {author} {\bibinfo {author} {\bibfnamefont {C.}~\bibnamefont
  {Wang}}\ and\ \bibinfo {author} {\bibfnamefont {J.~M.}\ \bibnamefont
  {Gertler}},\ }\bibfield  {title} {\bibinfo {title} {{Autonomous quantum state
  transfer by dissipation engineering}},\ }\href
  {https://doi.org/10.1103/PhysRevResearch.1.033198} {\bibfield  {journal}
  {\bibinfo  {journal} {Phys. Rev. Res.}\ }\textbf {\bibinfo {volume} {1}},\
  \bibinfo {pages} {033198} (\bibinfo {year} {2019})}\BibitemShut {NoStop}%
\bibitem [{\citenamefont {Liu}\ \emph {et~al.}(2011)\citenamefont {Liu},
  \citenamefont {Li}, \citenamefont {Huang}, \citenamefont {Li}, \citenamefont
  {Guo}, \citenamefont {Laine}, \citenamefont {Breuer},\ and\ \citenamefont
  {Piilo}}]{Liu2011}%
  \BibitemOpen
  \bibfield  {author} {\bibinfo {author} {\bibfnamefont {B.~H.}\ \bibnamefont
  {Liu}}, \bibinfo {author} {\bibfnamefont {L.}~\bibnamefont {Li}}, \bibinfo
  {author} {\bibfnamefont {Y.~F.}\ \bibnamefont {Huang}}, \bibinfo {author}
  {\bibfnamefont {C.~F.}\ \bibnamefont {Li}}, \bibinfo {author} {\bibfnamefont
  {G.~C.}\ \bibnamefont {Guo}}, \bibinfo {author} {\bibfnamefont {E.~M.}\
  \bibnamefont {Laine}}, \bibinfo {author} {\bibfnamefont {H.~P.}\ \bibnamefont
  {Breuer}},\ and\ \bibinfo {author} {\bibfnamefont {J.}~\bibnamefont
  {Piilo}},\ }\bibfield  {title} {\bibinfo {title} {{Experimental control of
  the transition from Markovian to non-Markovian dynamics of open quantum
  systems}},\ }\href {https://doi.org/10.1038/NPHYS2085} {\bibfield  {journal}
  {\bibinfo  {journal} {Nat. Phys.}\ }\textbf {\bibinfo {volume} {7}},\
  \bibinfo {pages} {931} (\bibinfo {year} {2011})}\BibitemShut {NoStop}%
\bibitem [{\citenamefont {Metelmann}\ and\ \citenamefont
  {Clerk}(2015)}]{Metelmann2015}%
  \BibitemOpen
  \bibfield  {author} {\bibinfo {author} {\bibfnamefont {A.}~\bibnamefont
  {Metelmann}}\ and\ \bibinfo {author} {\bibfnamefont {A.~A.}\ \bibnamefont
  {Clerk}},\ }\bibfield  {title} {\bibinfo {title} {{Nonreciprocal photon
  transmission and amplification via reservoir engineering}},\ }\href
  {https://doi.org/10.1103/PhysRevX.5.021025} {\bibfield  {journal} {\bibinfo
  {journal} {Phys. Rev. X}\ }\textbf {\bibinfo {volume} {5}},\ \bibinfo {pages}
  {021025} (\bibinfo {year} {2015})}\BibitemShut {NoStop}%
\bibitem [{\citenamefont {Abah}\ and\ \citenamefont {Lutz}(2014)}]{Abah2014}%
  \BibitemOpen
  \bibfield  {author} {\bibinfo {author} {\bibfnamefont {O.}~\bibnamefont
  {Abah}}\ and\ \bibinfo {author} {\bibfnamefont {E.}~\bibnamefont {Lutz}},\
  }\bibfield  {title} {\bibinfo {title} {{Efficiency of heat engines coupled to
  nonequilibrium reservoirs}},\ }\href
  {https://doi.org/10.1209/0295-5075/106/20001} {\bibfield  {journal} {\bibinfo
   {journal} {EPL}\ }\textbf {\bibinfo {volume} {106}},\ \bibinfo {pages}
  {20001} (\bibinfo {year} {2014})}\BibitemShut {NoStop}%
\bibitem [{\citenamefont {{De Assis}}\ \emph {et~al.}(2019)\citenamefont {{De
  Assis}}, \citenamefont {{De Mendon{\c{c}}a}}, \citenamefont {Villas-Boas},
  \citenamefont {{De Souza}}, \citenamefont {Sarthour}, \citenamefont
  {Oliveira},\ and\ \citenamefont {{De Almeida}}}]{DeAssis2019}%
  \BibitemOpen
  \bibfield  {author} {\bibinfo {author} {\bibfnamefont {R.~J.}\ \bibnamefont
  {{De Assis}}}, \bibinfo {author} {\bibfnamefont {T.~M.}\ \bibnamefont {{De
  Mendon{\c{c}}a}}}, \bibinfo {author} {\bibfnamefont {C.~J.}\ \bibnamefont
  {Villas-Boas}}, \bibinfo {author} {\bibfnamefont {A.~M.}\ \bibnamefont {{De
  Souza}}}, \bibinfo {author} {\bibfnamefont {R.~S.}\ \bibnamefont {Sarthour}},
  \bibinfo {author} {\bibfnamefont {I.~S.}\ \bibnamefont {Oliveira}},\ and\
  \bibinfo {author} {\bibfnamefont {N.~G.}\ \bibnamefont {{De Almeida}}},\
  }\bibfield  {title} {\bibinfo {title} {{Efficiency of a quantum Otto heat
  engine operating under a reservoir at effective negative temperatures}},\
  }\href {https://doi.org/10.1103/PhysRevLett.122.240602} {\bibfield  {journal}
  {\bibinfo  {journal} {Phys. Rev. Lett.}\ }\textbf {\bibinfo {volume} {122}},\
  \bibinfo {pages} {240602} (\bibinfo {year} {2019})}\BibitemShut {NoStop}%
\bibitem [{\citenamefont {Gardiner}\ and\ \citenamefont
  {Collett}(1985)}]{Gardiner1985}%
  \BibitemOpen
  \bibfield  {author} {\bibinfo {author} {\bibfnamefont {C.~W.}\ \bibnamefont
  {Gardiner}}\ and\ \bibinfo {author} {\bibfnamefont {M.~J.}\ \bibnamefont
  {Collett}},\ }\bibfield  {title} {\bibinfo {title} {{Input and output in
  damped quantum systems: Quantum stochastic differential equations and the
  master equation}},\ }\href {https://doi.org/10.1103/PhysRevA.31.3761}
  {\bibfield  {journal} {\bibinfo  {journal} {Phys. Rev. A}\ }\textbf {\bibinfo
  {volume} {31}},\ \bibinfo {pages} {3761} (\bibinfo {year}
  {1985})}\BibitemShut {NoStop}%
\bibitem [{\citenamefont {Gardiner}\ and\ \citenamefont
  {Zoller}(2004)}]{Gardiner2004}%
  \BibitemOpen
  \bibfield  {author} {\bibinfo {author} {\bibfnamefont {C.~W.}\ \bibnamefont
  {Gardiner}}\ and\ \bibinfo {author} {\bibfnamefont {P.}~\bibnamefont
  {Zoller}},\ }\href@noop {} {\emph {\bibinfo {title} {{Quantum noise: A
  handbook of Markovian and non-Markovian quantum stochastic methods with
  applications to quantum optics}}}}\ (\bibinfo  {publisher} {Springer,
  Berlin},\ \bibinfo {year} {2004})\BibitemShut {NoStop}%
\bibitem [{\citenamefont {Manzano}(2018)}]{Manzano2018}%
  \BibitemOpen
  \bibfield  {author} {\bibinfo {author} {\bibfnamefont {G.}~\bibnamefont
  {Manzano}},\ }\bibfield  {title} {\bibinfo {title} {{Squeezed thermal
  reservoir as a generalized equilibrium reservoir}},\ }\href
  {https://doi.org/10.1103/PhysRevE.98.042123} {\bibfield  {journal} {\bibinfo
  {journal} {Phys. Rev. E}\ }\textbf {\bibinfo {volume} {98}},\ \bibinfo
  {pages} {042123} (\bibinfo {year} {2018})}\BibitemShut {NoStop}%
\bibitem [{\citenamefont {Loudon}\ and\ \citenamefont
  {Knight}(1987)}]{Loudon1987}%
  \BibitemOpen
  \bibfield  {author} {\bibinfo {author} {\bibfnamefont {R.}~\bibnamefont
  {Loudon}}\ and\ \bibinfo {author} {\bibfnamefont {P.~L.}\ \bibnamefont
  {Knight}},\ }\bibfield  {title} {\bibinfo {title} {{Squeezed light}},\ }\href
  {https://doi.org/10.1080/09500348714550721} {\bibfield  {journal} {\bibinfo
  {journal} {J. Mod. Opt.}\ }\textbf {\bibinfo {volume} {34}},\ \bibinfo
  {pages} {709} (\bibinfo {year} {1987})}\BibitemShut {NoStop}%
\bibitem [{\citenamefont {Kitagawa}\ and\ \citenamefont
  {Ueda}(1993)}]{Kitagawa1993}%
  \BibitemOpen
  \bibfield  {author} {\bibinfo {author} {\bibfnamefont {M.}~\bibnamefont
  {Kitagawa}}\ and\ \bibinfo {author} {\bibfnamefont {M.}~\bibnamefont
  {Ueda}},\ }\bibfield  {title} {\bibinfo {title} {{Squeezed spin states}},\
  }\href@noop {} {\bibfield  {journal} {\bibinfo  {journal} {Phys. Rev. A}\
  }\textbf {\bibinfo {volume} {47}},\ \bibinfo {pages} {5138} (\bibinfo {year}
  {1993})}\BibitemShut {NoStop}%
\bibitem [{\citenamefont {Braunstein}\ and\ \citenamefont {{Van
  Loock}}(2005)}]{Braunstein2005}%
  \BibitemOpen
  \bibfield  {author} {\bibinfo {author} {\bibfnamefont {L.~S.}\ \bibnamefont
  {Braunstein}}\ and\ \bibinfo {author} {\bibfnamefont {P.}~\bibnamefont {{Van
  Loock}}},\ }\bibfield  {title} {\bibinfo {title} {{Quantum information with
  continuous variables}},\ }\href {https://doi.org/10.1103/RevModPhys.77.513}
  {\bibfield  {journal} {\bibinfo  {journal} {Rev. Mod. Phys.}\ }\textbf
  {\bibinfo {volume} {77}},\ \bibinfo {pages} {513} (\bibinfo {year}
  {2005})}\BibitemShut {NoStop}%
\bibitem [{\citenamefont {Madsen}\ \emph {et~al.}(2022)\citenamefont {Madsen},
  \citenamefont {Laudenbach}, \citenamefont {Askarani}, \citenamefont
  {Rortais}, \citenamefont {Vincent}, \citenamefont {Bulmer}, \citenamefont
  {Miatto}, \citenamefont {Neuhaus}, \citenamefont {Helt}, \citenamefont
  {Collins}, \citenamefont {Lita}, \citenamefont {Gerrits}, \citenamefont
  {Nam}, \citenamefont {Vaidya}, \citenamefont {Menotti}, \citenamefont
  {Dhand}, \citenamefont {Vernon}, \citenamefont {Quesada},\ and\ \citenamefont
  {Lavoie}}]{Madsen2022}%
  \BibitemOpen
  \bibfield  {author} {\bibinfo {author} {\bibfnamefont {L.~S.}\ \bibnamefont
  {Madsen}}, \bibinfo {author} {\bibfnamefont {F.}~\bibnamefont {Laudenbach}},
  \bibinfo {author} {\bibfnamefont {M.~F.}\ \bibnamefont {Askarani}}, \bibinfo
  {author} {\bibfnamefont {F.}~\bibnamefont {Rortais}}, \bibinfo {author}
  {\bibfnamefont {T.}~\bibnamefont {Vincent}}, \bibinfo {author} {\bibfnamefont
  {J.~F.}\ \bibnamefont {Bulmer}}, \bibinfo {author} {\bibfnamefont {F.~M.}\
  \bibnamefont {Miatto}}, \bibinfo {author} {\bibfnamefont {L.}~\bibnamefont
  {Neuhaus}}, \bibinfo {author} {\bibfnamefont {L.~G.}\ \bibnamefont {Helt}},
  \bibinfo {author} {\bibfnamefont {M.~J.}\ \bibnamefont {Collins}}, \bibinfo
  {author} {\bibfnamefont {A.~E.}\ \bibnamefont {Lita}}, \bibinfo {author}
  {\bibfnamefont {T.}~\bibnamefont {Gerrits}}, \bibinfo {author} {\bibfnamefont
  {S.~W.}\ \bibnamefont {Nam}}, \bibinfo {author} {\bibfnamefont {V.~D.}\
  \bibnamefont {Vaidya}}, \bibinfo {author} {\bibfnamefont {M.}~\bibnamefont
  {Menotti}}, \bibinfo {author} {\bibfnamefont {I.}~\bibnamefont {Dhand}},
  \bibinfo {author} {\bibfnamefont {Z.}~\bibnamefont {Vernon}}, \bibinfo
  {author} {\bibfnamefont {N.}~\bibnamefont {Quesada}},\ and\ \bibinfo {author}
  {\bibfnamefont {J.}~\bibnamefont {Lavoie}},\ }\bibfield  {title} {\bibinfo
  {title} {{Quantum computational advantage with a programmable photonic
  processor}},\ }\href {https://doi.org/10.1038/s41586-022-04725-x} {\bibfield
  {journal} {\bibinfo  {journal} {Nature (London)}\ }\textbf {\bibinfo {volume}
  {606}},\ \bibinfo {pages} {75} (\bibinfo {year} {2022})}\BibitemShut
  {NoStop}%
\bibitem [{\citenamefont {Bondurant}\ and\ \citenamefont
  {Shapiro}(1984)}]{Bondurant1984}%
  \BibitemOpen
  \bibfield  {author} {\bibinfo {author} {\bibfnamefont {R.~S.}\ \bibnamefont
  {Bondurant}}\ and\ \bibinfo {author} {\bibfnamefont {J.~H.}\ \bibnamefont
  {Shapiro}},\ }\bibfield  {title} {\bibinfo {title} {{Squeezed states in
  phase-sensing interferometers}},\ }\href
  {https://doi.org/10.1103/PhysRevD.30.2548} {\bibfield  {journal} {\bibinfo
  {journal} {Phys. Rev. D}\ }\textbf {\bibinfo {volume} {30}},\ \bibinfo
  {pages} {2548} (\bibinfo {year} {1984})}\BibitemShut {NoStop}%
\bibitem [{\citenamefont {Jia}\ and\ \citenamefont {Others}(2024)}]{Jia2024}%
  \BibitemOpen
  \bibfield  {author} {\bibinfo {author} {\bibfnamefont {W.}~\bibnamefont
  {Jia}}\ and\ \bibinfo {author} {\bibnamefont {Others}},\ }\bibfield  {title}
  {\bibinfo {title} {{Squeezing the quantum noise of a gravitational-wave
  detector below the standard quantum limit}},\ }\href
  {https://doi.org/10.1126/SCIENCE.ADO8069} {\bibfield  {journal} {\bibinfo
  {journal} {Science}\ }\textbf {\bibinfo {volume} {385}},\ \bibinfo {pages}
  {1318} (\bibinfo {year} {2024})}\BibitemShut {NoStop}%
\bibitem [{\citenamefont {Peano}\ \emph {et~al.}(2016)\citenamefont {Peano},
  \citenamefont {Houde}, \citenamefont {Brendel}, \citenamefont {Marquardt},\
  and\ \citenamefont {Clerk}}]{Peano2016}%
  \BibitemOpen
  \bibfield  {author} {\bibinfo {author} {\bibfnamefont {V.}~\bibnamefont
  {Peano}}, \bibinfo {author} {\bibfnamefont {M.}~\bibnamefont {Houde}},
  \bibinfo {author} {\bibfnamefont {C.}~\bibnamefont {Brendel}}, \bibinfo
  {author} {\bibfnamefont {F.}~\bibnamefont {Marquardt}},\ and\ \bibinfo
  {author} {\bibfnamefont {A.~A.}\ \bibnamefont {Clerk}},\ }\bibfield  {title}
  {\bibinfo {title} {{Topological phase transitions and chiral inelastic
  transport induced by the squeezing of light}},\ }\href
  {https://doi.org/10.1038/ncomms10779} {\bibfield  {journal} {\bibinfo
  {journal} {Nat. Commun.}\ }\textbf {\bibinfo {volume} {7}},\ \bibinfo {pages}
  {10779} (\bibinfo {year} {2016})}\BibitemShut {NoStop}%
\bibitem [{\citenamefont {Zhu}\ \emph {et~al.}(2020)\citenamefont {Zhu},
  \citenamefont {Ping}, \citenamefont {Yang},\ and\ \citenamefont
  {Agarwal}}]{Zhu2020}%
  \BibitemOpen
  \bibfield  {author} {\bibinfo {author} {\bibfnamefont {C.~J.}\ \bibnamefont
  {Zhu}}, \bibinfo {author} {\bibfnamefont {L.~L.}\ \bibnamefont {Ping}},
  \bibinfo {author} {\bibfnamefont {Y.~P.}\ \bibnamefont {Yang}},\ and\
  \bibinfo {author} {\bibfnamefont {G.~S.}\ \bibnamefont {Agarwal}},\
  }\bibfield  {title} {\bibinfo {title} {{Squeezed light induced symmetry
  breaking superradiant phase transition}},\ }\href
  {https://doi.org/10.1103/PhysRevLett.124.073602} {\bibfield  {journal}
  {\bibinfo  {journal} {Phys. Rev. Lett.}\ }\textbf {\bibinfo {volume} {124}},\
  \bibinfo {pages} {073602} (\bibinfo {year} {2020})}\BibitemShut {NoStop}%
\bibitem [{\citenamefont {Huang}\ \emph {et~al.}(2012)\citenamefont {Huang},
  \citenamefont {Wang},\ and\ \citenamefont {Yi}}]{Huang2012}%
  \BibitemOpen
  \bibfield  {author} {\bibinfo {author} {\bibfnamefont {X.~L.}\ \bibnamefont
  {Huang}}, \bibinfo {author} {\bibfnamefont {T.}~\bibnamefont {Wang}},\ and\
  \bibinfo {author} {\bibfnamefont {X.~X.}\ \bibnamefont {Yi}},\ }\bibfield
  {title} {\bibinfo {title} {{Effects of reservoir squeezing on quantum systems
  and work extraction}},\ }\href {https://doi.org/10.1103/PhysRevE.86.051105}
  {\bibfield  {journal} {\bibinfo  {journal} {Phys. Rev. E}\ }\textbf {\bibinfo
  {volume} {86}},\ \bibinfo {pages} {051105} (\bibinfo {year}
  {2012})}\BibitemShut {NoStop}%
\bibitem [{\citenamefont {Ro{\ss}nagel}\ \emph {et~al.}(2014)\citenamefont
  {Ro{\ss}nagel}, \citenamefont {Abah}, \citenamefont {Schmidt-Kaler},
  \citenamefont {Singer},\ and\ \citenamefont {Lutz}}]{Roßnagel2014}%
  \BibitemOpen
  \bibfield  {author} {\bibinfo {author} {\bibfnamefont {J.}~\bibnamefont
  {Ro{\ss}nagel}}, \bibinfo {author} {\bibfnamefont {O.}~\bibnamefont {Abah}},
  \bibinfo {author} {\bibfnamefont {F.}~\bibnamefont {Schmidt-Kaler}}, \bibinfo
  {author} {\bibfnamefont {K.}~\bibnamefont {Singer}},\ and\ \bibinfo {author}
  {\bibfnamefont {E.}~\bibnamefont {Lutz}},\ }\bibfield  {title} {\bibinfo
  {title} {{Nanoscale heat engine beyond the Carnot limit}},\ }\href
  {https://doi.org/10.1103/PhysRevLett.112.030602} {\bibfield  {journal}
  {\bibinfo  {journal} {Phys. Rev. Lett.}\ }\textbf {\bibinfo {volume} {112}},\
  \bibinfo {pages} {030602} (\bibinfo {year} {2014})}\BibitemShut {NoStop}%
\bibitem [{\citenamefont {Manzano}\ \emph {et~al.}(2016)\citenamefont
  {Manzano}, \citenamefont {Galve}, \citenamefont {Zambrini},\ and\
  \citenamefont {Parrondo}}]{Manzano2016}%
  \BibitemOpen
  \bibfield  {author} {\bibinfo {author} {\bibfnamefont {G.}~\bibnamefont
  {Manzano}}, \bibinfo {author} {\bibfnamefont {F.}~\bibnamefont {Galve}},
  \bibinfo {author} {\bibfnamefont {R.}~\bibnamefont {Zambrini}},\ and\
  \bibinfo {author} {\bibfnamefont {J.~M.}\ \bibnamefont {Parrondo}},\
  }\bibfield  {title} {\bibinfo {title} {{Entropy production and thermodynamic
  power of the squeezed thermal reservoir}},\ }\href
  {https://doi.org/10.1103/PhysRevE.93.052120} {\bibfield  {journal} {\bibinfo
  {journal} {Phys. Rev. E}\ }\textbf {\bibinfo {volume} {93}},\ \bibinfo
  {pages} {052120} (\bibinfo {year} {2016})}\BibitemShut {NoStop}%
\bibitem [{\citenamefont {Klaers}\ \emph {et~al.}(2017)\citenamefont {Klaers},
  \citenamefont {Faelt}, \citenamefont {Imamoglu},\ and\ \citenamefont
  {Togan}}]{Klaers2017}%
  \BibitemOpen
  \bibfield  {author} {\bibinfo {author} {\bibfnamefont {J.}~\bibnamefont
  {Klaers}}, \bibinfo {author} {\bibfnamefont {S.}~\bibnamefont {Faelt}},
  \bibinfo {author} {\bibfnamefont {A.}~\bibnamefont {Imamoglu}},\ and\
  \bibinfo {author} {\bibfnamefont {E.}~\bibnamefont {Togan}},\ }\bibfield
  {title} {\bibinfo {title} {{Squeezed thermal reservoirs as a resource for a
  nanomechanical engine beyond the Carnot limit}},\ }\href
  {https://doi.org/10.1103/PhysRevX.7.031044} {\bibfield  {journal} {\bibinfo
  {journal} {Phys. Rev. X}\ }\textbf {\bibinfo {volume} {7}},\ \bibinfo {pages}
  {031044} (\bibinfo {year} {2017})}\BibitemShut {NoStop}%
\bibitem [{\citenamefont {Gardiner}(1986)}]{Gardiner1986}%
  \BibitemOpen
  \bibfield  {author} {\bibinfo {author} {\bibfnamefont {C.~W.}\ \bibnamefont
  {Gardiner}},\ }\bibfield  {title} {\bibinfo {title} {{Inhibition of atomic
  phase decays by squeezed light: A direct effect of squeezing}},\ }\href
  {https://doi.org/10.1103/PhysRevLett.56.1917} {\bibfield  {journal} {\bibinfo
   {journal} {Phys. Rev. Lett.}\ }\textbf {\bibinfo {volume} {56}},\ \bibinfo
  {pages} {1917} (\bibinfo {year} {1986})}\BibitemShut {NoStop}%
\bibitem [{\citenamefont {Murch}\ \emph {et~al.}(2013)\citenamefont {Murch},
  \citenamefont {Weber}, \citenamefont {Beck}, \citenamefont {Ginossar},\ and\
  \citenamefont {Siddiqi}}]{Murch2013}%
  \BibitemOpen
  \bibfield  {author} {\bibinfo {author} {\bibfnamefont {K.~W.}\ \bibnamefont
  {Murch}}, \bibinfo {author} {\bibfnamefont {S.~J.}\ \bibnamefont {Weber}},
  \bibinfo {author} {\bibfnamefont {K.~M.}\ \bibnamefont {Beck}}, \bibinfo
  {author} {\bibfnamefont {E.}~\bibnamefont {Ginossar}},\ and\ \bibinfo
  {author} {\bibfnamefont {I.}~\bibnamefont {Siddiqi}},\ }\bibfield  {title}
  {\bibinfo {title} {{Reduction of the radiative decay of atomic coherence in
  squeezed vacuum}},\ }\href {https://doi.org/10.1038/nature12264} {\bibfield
  {journal} {\bibinfo  {journal} {Nature (London)}\ }\textbf {\bibinfo {volume}
  {499}},\ \bibinfo {pages} {62} (\bibinfo {year} {2013})}\BibitemShut
  {NoStop}%
\bibitem [{\citenamefont {Swain}(1994)}]{Swain1994}%
  \BibitemOpen
  \bibfield  {author} {\bibinfo {author} {\bibfnamefont {S.}~\bibnamefont
  {Swain}},\ }\bibfield  {title} {\bibinfo {title} {{Anomalous resonance
  fluorescence spectra in a squeezed vacuum}},\ }\href
  {https://doi.org/10.1103/PhysRevLett.73.1493} {\bibfield  {journal} {\bibinfo
   {journal} {Phys. Rev. Lett.}\ }\textbf {\bibinfo {volume} {73}},\ \bibinfo
  {pages} {1493} (\bibinfo {year} {1994})}\BibitemShut {NoStop}%
\bibitem [{\citenamefont {Toyli}\ \emph {et~al.}(2016)\citenamefont {Toyli},
  \citenamefont {Eddins}, \citenamefont {Boutin}, \citenamefont {Puri},
  \citenamefont {Hover}, \citenamefont {Bolkhovsky}, \citenamefont {Oliver},
  \citenamefont {Blais},\ and\ \citenamefont {Siddiqi}}]{Toyli2016}%
  \BibitemOpen
  \bibfield  {author} {\bibinfo {author} {\bibfnamefont {D.~M.}\ \bibnamefont
  {Toyli}}, \bibinfo {author} {\bibfnamefont {A.~W.}\ \bibnamefont {Eddins}},
  \bibinfo {author} {\bibfnamefont {S.}~\bibnamefont {Boutin}}, \bibinfo
  {author} {\bibfnamefont {S.}~\bibnamefont {Puri}}, \bibinfo {author}
  {\bibfnamefont {D.}~\bibnamefont {Hover}}, \bibinfo {author} {\bibfnamefont
  {V.}~\bibnamefont {Bolkhovsky}}, \bibinfo {author} {\bibfnamefont {W.~D.}\
  \bibnamefont {Oliver}}, \bibinfo {author} {\bibfnamefont {A.}~\bibnamefont
  {Blais}},\ and\ \bibinfo {author} {\bibfnamefont {I.}~\bibnamefont
  {Siddiqi}},\ }\bibfield  {title} {\bibinfo {title} {{Resonance fluorescence
  from an artificial atom in squeezed vacuum}},\ }\href
  {https://doi.org/10.1103/PhysRevX.6.031004} {\bibfield  {journal} {\bibinfo
  {journal} {Phys. Rev. X}\ }\textbf {\bibinfo {volume} {6}},\ \bibinfo {pages}
  {031004} (\bibinfo {year} {2016})}\BibitemShut {NoStop}%
\bibitem [{\citenamefont {Parkins}\ \emph {et~al.}(2006)\citenamefont
  {Parkins}, \citenamefont {Solano},\ and\ \citenamefont
  {Cirac}}]{Parkins2006}%
  \BibitemOpen
  \bibfield  {author} {\bibinfo {author} {\bibfnamefont {A.~S.}\ \bibnamefont
  {Parkins}}, \bibinfo {author} {\bibfnamefont {E.}~\bibnamefont {Solano}},\
  and\ \bibinfo {author} {\bibfnamefont {J.~I.}\ \bibnamefont {Cirac}},\
  }\bibfield  {title} {\bibinfo {title} {{Unconditional two-mode squeezing of
  separated atomic ensembles}},\ }\href
  {https://doi.org/10.1103/PhysRevLett.96.053602} {\bibfield  {journal}
  {\bibinfo  {journal} {Phys. Rev. Lett.}\ }\textbf {\bibinfo {volume} {96}},\
  \bibinfo {pages} {053602} (\bibinfo {year} {2006})}\BibitemShut {NoStop}%
\bibitem [{\citenamefont {Porras}\ and\ \citenamefont
  {Garc{\'{i}}a-Ripoll}(2012)}]{Porras2012}%
  \BibitemOpen
  \bibfield  {author} {\bibinfo {author} {\bibfnamefont {D.}~\bibnamefont
  {Porras}}\ and\ \bibinfo {author} {\bibfnamefont {J.~J.}\ \bibnamefont
  {Garc{\'{i}}a-Ripoll}},\ }\bibfield  {title} {\bibinfo {title} {{Shaping an
  itinerant quantum field into a multimode squeezed vacuum by dissipation}},\
  }\href {https://doi.org/10.1103/PhysRevLett.108.043602} {\bibfield  {journal}
  {\bibinfo  {journal} {Phys. Rev. Lett.}\ }\textbf {\bibinfo {volume} {108}},\
  \bibinfo {pages} {043602} (\bibinfo {year} {2012})}\BibitemShut {NoStop}%
\bibitem [{\citenamefont {Kronwald}\ \emph {et~al.}(2014)\citenamefont
  {Kronwald}, \citenamefont {Marquardt},\ and\ \citenamefont
  {Clerk}}]{Kronwald2014}%
  \BibitemOpen
  \bibfield  {author} {\bibinfo {author} {\bibfnamefont {A.}~\bibnamefont
  {Kronwald}}, \bibinfo {author} {\bibfnamefont {F.}~\bibnamefont
  {Marquardt}},\ and\ \bibinfo {author} {\bibfnamefont {A.~A.}\ \bibnamefont
  {Clerk}},\ }\bibfield  {title} {\bibinfo {title} {{Dissipative optomechanical
  squeezing of light}},\ }\href {https://doi.org/10.1088/1367-2630/16/6/063058}
  {\bibfield  {journal} {\bibinfo  {journal} {New J. Phys.}\ }\textbf {\bibinfo
  {volume} {16}},\ \bibinfo {pages} {063058} (\bibinfo {year}
  {2014})}\BibitemShut {NoStop}%
\bibitem [{\citenamefont {Wollman}\ \emph {et~al.}(2015)\citenamefont
  {Wollman}, \citenamefont {Lei}, \citenamefont {Weinstein}, \citenamefont
  {Suh}, \citenamefont {Kronwald}, \citenamefont {Marquardt}, \citenamefont
  {Clerk},\ and\ \citenamefont {Schwab}}]{Wollman2015}%
  \BibitemOpen
  \bibfield  {author} {\bibinfo {author} {\bibfnamefont {E.~E.}\ \bibnamefont
  {Wollman}}, \bibinfo {author} {\bibfnamefont {C.~U.}\ \bibnamefont {Lei}},
  \bibinfo {author} {\bibfnamefont {A.~J.}\ \bibnamefont {Weinstein}}, \bibinfo
  {author} {\bibfnamefont {J.}~\bibnamefont {Suh}}, \bibinfo {author}
  {\bibfnamefont {A.}~\bibnamefont {Kronwald}}, \bibinfo {author}
  {\bibfnamefont {F.}~\bibnamefont {Marquardt}}, \bibinfo {author}
  {\bibfnamefont {A.~A.}\ \bibnamefont {Clerk}},\ and\ \bibinfo {author}
  {\bibfnamefont {K.~C.}\ \bibnamefont {Schwab}},\ }\bibfield  {title}
  {\bibinfo {title} {{Quantum squeezing of motion in a mechanical resonator}},\
  }\href {https://doi.org/10.1126/SCIENCE.AAC5138} {\bibfield  {journal}
  {\bibinfo  {journal} {Science}\ }\textbf {\bibinfo {volume} {349}},\ \bibinfo
  {pages} {952} (\bibinfo {year} {2015})}\BibitemShut {NoStop}%
\bibitem [{\citenamefont {Bai}\ and\ \citenamefont {An}(2021)}]{Bai2021}%
  \BibitemOpen
  \bibfield  {author} {\bibinfo {author} {\bibfnamefont {S.~Y.}\ \bibnamefont
  {Bai}}\ and\ \bibinfo {author} {\bibfnamefont {J.~H.}\ \bibnamefont {An}},\
  }\bibfield  {title} {\bibinfo {title} {{Generating stable spin squeezing by
  squeezed-reservoir engineering}},\ }\href
  {https://doi.org/10.1103/PhysRevLett.127.083602} {\bibfield  {journal}
  {\bibinfo  {journal} {Phys. Rev. Lett.}\ }\textbf {\bibinfo {volume} {127}},\
  \bibinfo {pages} {083602} (\bibinfo {year} {2021})}\BibitemShut {NoStop}%
\bibitem [{\citenamefont {Kraus}\ and\ \citenamefont
  {Cirac}(2004)}]{Kraus2004}%
  \BibitemOpen
  \bibfield  {author} {\bibinfo {author} {\bibfnamefont {B.}~\bibnamefont
  {Kraus}}\ and\ \bibinfo {author} {\bibfnamefont {J.~I.}\ \bibnamefont
  {Cirac}},\ }\bibfield  {title} {\bibinfo {title} {{Discrete entanglement
  distribution with squeezed light}},\ }\href
  {https://doi.org/10.1103/PhysRevLett.92.013602} {\bibfield  {journal}
  {\bibinfo  {journal} {Phys. Rev. Lett.}\ }\textbf {\bibinfo {volume} {92}},\
  \bibinfo {pages} {013602} (\bibinfo {year} {2004})}\BibitemShut {NoStop}%
\bibitem [{\citenamefont {Banerjee}\ \emph {et~al.}(2010)\citenamefont
  {Banerjee}, \citenamefont {Ravishankar},\ and\ \citenamefont
  {Srikanth}}]{Banerjee2010}%
  \BibitemOpen
  \bibfield  {author} {\bibinfo {author} {\bibfnamefont {S.}~\bibnamefont
  {Banerjee}}, \bibinfo {author} {\bibfnamefont {V.}~\bibnamefont
  {Ravishankar}},\ and\ \bibinfo {author} {\bibfnamefont {R.}~\bibnamefont
  {Srikanth}},\ }\bibfield  {title} {\bibinfo {title} {{Dynamics of
  entanglement in two-qubit open system interacting with a squeezed thermal
  bath via dissipative interaction}},\ }\href
  {https://doi.org/10.1016/j.aop.2010.01.003} {\bibfield  {journal} {\bibinfo
  {journal} {Ann. Phys. (N. Y).}\ }\textbf {\bibinfo {volume} {325}},\ \bibinfo
  {pages} {816} (\bibinfo {year} {2010})}\BibitemShut {NoStop}%
\bibitem [{\citenamefont {Wang}\ and\ \citenamefont {Clerk}(2013)}]{Wang2013}%
  \BibitemOpen
  \bibfield  {author} {\bibinfo {author} {\bibfnamefont {Y.~D.}\ \bibnamefont
  {Wang}}\ and\ \bibinfo {author} {\bibfnamefont {A.~A.}\ \bibnamefont
  {Clerk}},\ }\bibfield  {title} {\bibinfo {title} {{Reservoir-engineered
  entanglement in optomechanical systems}},\ }\href
  {https://doi.org/10.1103/PhysRevLett.110.253601} {\bibfield  {journal}
  {\bibinfo  {journal} {Phys. Rev. Lett.}\ }\textbf {\bibinfo {volume} {110}},\
  \bibinfo {pages} {253601} (\bibinfo {year} {2013})}\BibitemShut {NoStop}%
\bibitem [{\citenamefont {Zippilli}\ \emph {et~al.}(2015)\citenamefont
  {Zippilli}, \citenamefont {Li},\ and\ \citenamefont {Vitali}}]{Zippilli2015}%
  \BibitemOpen
  \bibfield  {author} {\bibinfo {author} {\bibfnamefont {S.}~\bibnamefont
  {Zippilli}}, \bibinfo {author} {\bibfnamefont {J.}~\bibnamefont {Li}},\ and\
  \bibinfo {author} {\bibfnamefont {D.}~\bibnamefont {Vitali}},\ }\bibfield
  {title} {\bibinfo {title} {{Steady-state nested entanglement structures in
  harmonic chains with single-site squeezing manipulation}},\ }\href
  {https://doi.org/10.1103/PhysRevA.92.032319} {\bibfield  {journal} {\bibinfo
  {journal} {Phys. Rev. A}\ }\textbf {\bibinfo {volume} {92}},\ \bibinfo
  {pages} {032319} (\bibinfo {year} {2015})}\BibitemShut {NoStop}%
\bibitem [{\citenamefont {Govia}\ \emph {et~al.}(2022)\citenamefont {Govia},
  \citenamefont {Lingenfelter},\ and\ \citenamefont {Clerk}}]{Govia2022}%
  \BibitemOpen
  \bibfield  {author} {\bibinfo {author} {\bibfnamefont {L.~C.}\ \bibnamefont
  {Govia}}, \bibinfo {author} {\bibfnamefont {A.}~\bibnamefont
  {Lingenfelter}},\ and\ \bibinfo {author} {\bibfnamefont {A.~A.}\ \bibnamefont
  {Clerk}},\ }\bibfield  {title} {\bibinfo {title} {{Stabilizing two-qubit
  entanglement by mimicking a squeezed environment}},\ }\href
  {https://doi.org/10.1103/PhysRevResearch.4.023010} {\bibfield  {journal}
  {\bibinfo  {journal} {Phys. Rev. Res.}\ }\textbf {\bibinfo {volume} {4}},\
  \bibinfo {pages} {023010} (\bibinfo {year} {2022})}\BibitemShut {NoStop}%
\bibitem [{\citenamefont {Kronwald}\ \emph {et~al.}(2013)\citenamefont
  {Kronwald}, \citenamefont {Marquardt},\ and\ \citenamefont
  {Clerk}}]{Kronwald2013}%
  \BibitemOpen
  \bibfield  {author} {\bibinfo {author} {\bibfnamefont {A.}~\bibnamefont
  {Kronwald}}, \bibinfo {author} {\bibfnamefont {F.}~\bibnamefont
  {Marquardt}},\ and\ \bibinfo {author} {\bibfnamefont {A.~A.}\ \bibnamefont
  {Clerk}},\ }\bibfield  {title} {\bibinfo {title} {{Arbitrarily large
  steady-state bosonic squeezing via dissipation}},\ }\href
  {https://doi.org/10.1103/PhysRevA.88.063833} {\bibfield  {journal} {\bibinfo
  {journal} {Phys. Rev. A}\ }\textbf {\bibinfo {volume} {88}},\ \bibinfo
  {pages} {063833} (\bibinfo {year} {2013})}\BibitemShut {NoStop}%
\bibitem [{\citenamefont {Shahmoon}\ and\ \citenamefont
  {Kurizki}(2013)}]{Shahmoon2013}%
  \BibitemOpen
  \bibfield  {author} {\bibinfo {author} {\bibfnamefont {E.}~\bibnamefont
  {Shahmoon}}\ and\ \bibinfo {author} {\bibfnamefont {G.}~\bibnamefont
  {Kurizki}},\ }\bibfield  {title} {\bibinfo {title} {{Engineering a thermal
  squeezed reservoir by energy-level modulation}},\ }\href
  {https://doi.org/10.1103/PhysRevA.87.013841} {\bibfield  {journal} {\bibinfo
  {journal} {Phys. Rev. A}\ }\textbf {\bibinfo {volume} {87}},\ \bibinfo
  {pages} {13841} (\bibinfo {year} {2013})}\BibitemShut {NoStop}%
\bibitem [{\citenamefont {{Breuer, Heinz-Peter and
  Petruccione}}(2002)}]{Breuer2002}%
  \BibitemOpen
  \bibfield  {author} {\bibinfo {author} {\bibfnamefont {F.}~\bibnamefont
  {{Breuer, Heinz-Peter and Petruccione}}},\ }\href@noop {} {\emph {\bibinfo
  {title} {{The theory of open quantum systems}}}}\ (\bibinfo  {publisher}
  {Oxford University Press, New York},\ \bibinfo {year} {2002})\BibitemShut
  {NoStop}%
\bibitem [{\citenamefont {J{\"{a}}ger}\ \emph {et~al.}(2022)\citenamefont
  {J{\"{a}}ger}, \citenamefont {Schmit}, \citenamefont {Morigi}, \citenamefont
  {Holland},\ and\ \citenamefont {Betzholz}}]{Jager2022}%
  \BibitemOpen
  \bibfield  {author} {\bibinfo {author} {\bibfnamefont {S.~B.}\ \bibnamefont
  {J{\"{a}}ger}}, \bibinfo {author} {\bibfnamefont {T.}~\bibnamefont {Schmit}},
  \bibinfo {author} {\bibfnamefont {G.}~\bibnamefont {Morigi}}, \bibinfo
  {author} {\bibfnamefont {M.~J.}\ \bibnamefont {Holland}},\ and\ \bibinfo
  {author} {\bibfnamefont {R.}~\bibnamefont {Betzholz}},\ }\bibfield  {title}
  {\bibinfo {title} {{Lindblad master equations for quantum systems coupled to
  dissipative bosonic modes}},\ }\href
  {https://doi.org/10.1103/PHYSREVLETT.129.063601} {\bibfield  {journal}
  {\bibinfo  {journal} {Phys. Rev. Lett.}\ }\textbf {\bibinfo {volume} {129}},\
  \bibinfo {pages} {063601} (\bibinfo {year} {2022})}\BibitemShut {NoStop}%
\bibitem [{\citenamefont {J{\"{a}}ger}\ and\ \citenamefont
  {Betzholz}(2023)}]{Jager2023}%
  \BibitemOpen
  \bibfield  {author} {\bibinfo {author} {\bibfnamefont {S.~B.}\ \bibnamefont
  {J{\"{a}}ger}}\ and\ \bibinfo {author} {\bibfnamefont {R.}~\bibnamefont
  {Betzholz}},\ }\bibfield  {title} {\bibinfo {title} {{Effective description
  of cooling and thermal shifts in quantum systems coupled to bosonic modes}},\
  }\href {https://doi.org/10.1103/PhysRevA.108.013110} {\bibfield  {journal}
  {\bibinfo  {journal} {Phys. Rev. A}\ }\textbf {\bibinfo {volume} {108}},\
  \bibinfo {pages} {013110} (\bibinfo {year} {2023})}\BibitemShut {NoStop}%
\bibitem [{\citenamefont {Haroche}\ and\ \citenamefont
  {Kleppner}(1989)}]{Haroche1989}%
  \BibitemOpen
  \bibfield  {author} {\bibinfo {author} {\bibfnamefont {S.}~\bibnamefont
  {Haroche}}\ and\ \bibinfo {author} {\bibfnamefont {D.}~\bibnamefont
  {Kleppner}},\ }\bibfield  {title} {\bibinfo {title} {{Cavity quantum
  electrodynamics}},\ }\href {https://doi.org/10.1063/1.881201} {\bibfield
  {journal} {\bibinfo  {journal} {Phys. Today}\ }\textbf {\bibinfo {volume}
  {42}},\ \bibinfo {pages} {24} (\bibinfo {year} {1989})}\BibitemShut {NoStop}%
\bibitem [{\citenamefont {Raimond}\ \emph {et~al.}(2001)\citenamefont
  {Raimond}, \citenamefont {Brune},\ and\ \citenamefont
  {Haroche}}]{Raimond2001}%
  \BibitemOpen
  \bibfield  {author} {\bibinfo {author} {\bibfnamefont {J.~M.}\ \bibnamefont
  {Raimond}}, \bibinfo {author} {\bibfnamefont {M.}~\bibnamefont {Brune}},\
  and\ \bibinfo {author} {\bibfnamefont {S.}~\bibnamefont {Haroche}},\
  }\bibfield  {title} {\bibinfo {title} {{Colloquium: Manipulating quantum
  entanglement with atoms and photons in a cavity}},\ }\href
  {https://doi.org/10.1103/RevModPhys.73.565} {\bibfield  {journal} {\bibinfo
  {journal} {Rev. Mod. Phys.}\ }\textbf {\bibinfo {volume} {73}},\ \bibinfo
  {pages} {565} (\bibinfo {year} {2001})}\BibitemShut {NoStop}%
\bibitem [{\citenamefont {Blais}\ \emph {et~al.}(2021)\citenamefont {Blais},
  \citenamefont {Grimsmo}, \citenamefont {Girvin},\ and\ \citenamefont
  {Wallraff}}]{Blais2021a}%
  \BibitemOpen
  \bibfield  {author} {\bibinfo {author} {\bibfnamefont {A.}~\bibnamefont
  {Blais}}, \bibinfo {author} {\bibfnamefont {A.~L.}\ \bibnamefont {Grimsmo}},
  \bibinfo {author} {\bibfnamefont {S.~M.}\ \bibnamefont {Girvin}},\ and\
  \bibinfo {author} {\bibfnamefont {A.}~\bibnamefont {Wallraff}},\ }\bibfield
  {title} {\bibinfo {title} {{Circuit quantum electrodynamics}},\ }\href
  {https://doi.org/10.1103/RevModPhys.93.025005} {\bibfield  {journal}
  {\bibinfo  {journal} {Rev. Mod. Phys.}\ }\textbf {\bibinfo {volume} {93}},\
  \bibinfo {pages} {025005} (\bibinfo {year} {2021})}\BibitemShut {NoStop}%
\bibitem [{\citenamefont {Wallraff}\ \emph {et~al.}(2004)\citenamefont
  {Wallraff}, \citenamefont {Schuster}, \citenamefont {Blais}, \citenamefont
  {Frunzio}, \citenamefont {Huang}, \citenamefont {Majer}, \citenamefont
  {Kumar}, \citenamefont {Girvin},\ and\ \citenamefont
  {Schoelkopf}}]{Wallraff2004}%
  \BibitemOpen
  \bibfield  {author} {\bibinfo {author} {\bibfnamefont {A.}~\bibnamefont
  {Wallraff}}, \bibinfo {author} {\bibfnamefont {D.~I.}\ \bibnamefont
  {Schuster}}, \bibinfo {author} {\bibfnamefont {A.}~\bibnamefont {Blais}},
  \bibinfo {author} {\bibfnamefont {L.}~\bibnamefont {Frunzio}}, \bibinfo
  {author} {\bibfnamefont {R.~S.}\ \bibnamefont {Huang}}, \bibinfo {author}
  {\bibfnamefont {J.}~\bibnamefont {Majer}}, \bibinfo {author} {\bibfnamefont
  {S.}~\bibnamefont {Kumar}}, \bibinfo {author} {\bibfnamefont {S.~M.}\
  \bibnamefont {Girvin}},\ and\ \bibinfo {author} {\bibfnamefont {R.~J.}\
  \bibnamefont {Schoelkopf}},\ }\bibfield  {title} {\bibinfo {title} {{Strong
  coupling of a single photon to a superconducting qubit using circuit quantum
  electrodynamics}},\ }\href {https://doi.org/10.1038/nature02851} {\bibfield
  {journal} {\bibinfo  {journal} {Nature (London)}\ }\textbf {\bibinfo {volume}
  {431}},\ \bibinfo {pages} {162} (\bibinfo {year} {2004})}\BibitemShut
  {NoStop}%
\bibitem [{\citenamefont {Leibfried}\ \emph {et~al.}(2003)\citenamefont
  {Leibfried}, \citenamefont {Blatt}, \citenamefont {Monroe},\ and\
  \citenamefont {Wineland}}]{Leibfried2003}%
  \BibitemOpen
  \bibfield  {author} {\bibinfo {author} {\bibfnamefont {D.}~\bibnamefont
  {Leibfried}}, \bibinfo {author} {\bibfnamefont {R.}~\bibnamefont {Blatt}},
  \bibinfo {author} {\bibfnamefont {C.}~\bibnamefont {Monroe}},\ and\ \bibinfo
  {author} {\bibfnamefont {D.}~\bibnamefont {Wineland}},\ }\bibfield  {title}
  {\bibinfo {title} {{Quantum dynamics of single trapped ions}},\ }\href
  {https://doi.org/10.1103/RevModPhys.75.281} {\bibfield  {journal} {\bibinfo
  {journal} {Rev. Mod. Phys.}\ }\textbf {\bibinfo {volume} {75}},\ \bibinfo
  {pages} {281} (\bibinfo {year} {2003})}\BibitemShut {NoStop}%
\bibitem [{\citenamefont {Lv}\ \emph {et~al.}(2018)\citenamefont {Lv},
  \citenamefont {An}, \citenamefont {Liu}, \citenamefont {Zhang}, \citenamefont
  {Pedernales}, \citenamefont {Lamata}, \citenamefont {Solano},\ and\
  \citenamefont {Kim}}]{Lv2018}%
  \BibitemOpen
  \bibfield  {author} {\bibinfo {author} {\bibfnamefont {D.}~\bibnamefont
  {Lv}}, \bibinfo {author} {\bibfnamefont {S.}~\bibnamefont {An}}, \bibinfo
  {author} {\bibfnamefont {Z.}~\bibnamefont {Liu}}, \bibinfo {author}
  {\bibfnamefont {J.~N.}\ \bibnamefont {Zhang}}, \bibinfo {author}
  {\bibfnamefont {J.~S.}\ \bibnamefont {Pedernales}}, \bibinfo {author}
  {\bibfnamefont {L.}~\bibnamefont {Lamata}}, \bibinfo {author} {\bibfnamefont
  {E.}~\bibnamefont {Solano}},\ and\ \bibinfo {author} {\bibfnamefont
  {K.}~\bibnamefont {Kim}},\ }\bibfield  {title} {\bibinfo {title} {{Quantum
  simulation of the quantum Rabi model in a trapped ion}},\ }\href
  {https://doi.org/10.1103/PhysRevX.8.021027} {\bibfield  {journal} {\bibinfo
  {journal} {Phys. Rev. X}\ }\textbf {\bibinfo {volume} {8}},\ \bibinfo {pages}
  {021027} (\bibinfo {year} {2018})}\BibitemShut {NoStop}%
\bibitem [{\citenamefont {Manucharyan}\ \emph {et~al.}(2009)\citenamefont
  {Manucharyan}, \citenamefont {Koch}, \citenamefont {Glazman},\ and\
  \citenamefont {Devoret}}]{Manucharyan2009}%
  \BibitemOpen
  \bibfield  {author} {\bibinfo {author} {\bibfnamefont {V.~E.}\ \bibnamefont
  {Manucharyan}}, \bibinfo {author} {\bibfnamefont {J.}~\bibnamefont {Koch}},
  \bibinfo {author} {\bibfnamefont {L.~I.}\ \bibnamefont {Glazman}},\ and\
  \bibinfo {author} {\bibfnamefont {M.~H.}\ \bibnamefont {Devoret}},\
  }\bibfield  {title} {\bibinfo {title} {{Fluxonium: Single cooper-pair circuit
  free of charge offsets}},\ }\href
  {https://doi.org/10.1126/SCIENCE.1175552/SUPPL_FILE/MANUCHARYAN.SOM.PDF}
  {\bibfield  {journal} {\bibinfo  {journal} {Science}\ }\textbf {\bibinfo
  {volume} {326}},\ \bibinfo {pages} {113} (\bibinfo {year}
  {2009})}\BibitemShut {NoStop}%
\bibitem [{\citenamefont {Nguyen}\ \emph {et~al.}(2019)\citenamefont {Nguyen},
  \citenamefont {Lin}, \citenamefont {Somoroff}, \citenamefont {Mencia},
  \citenamefont {Grabon},\ and\ \citenamefont {Manucharyan}}]{Nguyen2019}%
  \BibitemOpen
  \bibfield  {author} {\bibinfo {author} {\bibfnamefont {L.~B.}\ \bibnamefont
  {Nguyen}}, \bibinfo {author} {\bibfnamefont {Y.~H.}\ \bibnamefont {Lin}},
  \bibinfo {author} {\bibfnamefont {A.}~\bibnamefont {Somoroff}}, \bibinfo
  {author} {\bibfnamefont {R.}~\bibnamefont {Mencia}}, \bibinfo {author}
  {\bibfnamefont {N.}~\bibnamefont {Grabon}},\ and\ \bibinfo {author}
  {\bibfnamefont {V.~E.}\ \bibnamefont {Manucharyan}},\ }\bibfield  {title}
  {\bibinfo {title} {{High-coherence fluxonium qubit}},\ }\href
  {https://doi.org/10.1103/PHYSREVX.9.041041} {\bibfield  {journal} {\bibinfo
  {journal} {Phys. Rev. X}\ }\textbf {\bibinfo {volume} {9}},\ \bibinfo {pages}
  {041041} (\bibinfo {year} {2019})}\BibitemShut {NoStop}%
\bibitem [{\citenamefont {Nguyen}\ \emph {et~al.}(2022)\citenamefont {Nguyen},
  \citenamefont {Koolstra}, \citenamefont {Kim}, \citenamefont {Morvan},
  \citenamefont {Chistolini}, \citenamefont {Singh}, \citenamefont {Nesterov},
  \citenamefont {J{\"{u}}nger}, \citenamefont {Chen}, \citenamefont
  {Pedramrazi}, \citenamefont {Mitchell}, \citenamefont {Kreikebaum},
  \citenamefont {Puri}, \citenamefont {Santiago},\ and\ \citenamefont
  {Siddiqi}}]{Nguyen2022}%
  \BibitemOpen
  \bibfield  {author} {\bibinfo {author} {\bibfnamefont {L.~B.}\ \bibnamefont
  {Nguyen}}, \bibinfo {author} {\bibfnamefont {G.}~\bibnamefont {Koolstra}},
  \bibinfo {author} {\bibfnamefont {Y.}~\bibnamefont {Kim}}, \bibinfo {author}
  {\bibfnamefont {A.}~\bibnamefont {Morvan}}, \bibinfo {author} {\bibfnamefont
  {T.}~\bibnamefont {Chistolini}}, \bibinfo {author} {\bibfnamefont
  {S.}~\bibnamefont {Singh}}, \bibinfo {author} {\bibfnamefont {K.~N.}\
  \bibnamefont {Nesterov}}, \bibinfo {author} {\bibfnamefont {C.}~\bibnamefont
  {J{\"{u}}nger}}, \bibinfo {author} {\bibfnamefont {L.}~\bibnamefont {Chen}},
  \bibinfo {author} {\bibfnamefont {Z.}~\bibnamefont {Pedramrazi}}, \bibinfo
  {author} {\bibfnamefont {B.~K.}\ \bibnamefont {Mitchell}}, \bibinfo {author}
  {\bibfnamefont {J.~M.}\ \bibnamefont {Kreikebaum}}, \bibinfo {author}
  {\bibfnamefont {S.}~\bibnamefont {Puri}}, \bibinfo {author} {\bibfnamefont
  {D.~I.}\ \bibnamefont {Santiago}},\ and\ \bibinfo {author} {\bibfnamefont
  {I.}~\bibnamefont {Siddiqi}},\ }\bibfield  {title} {\bibinfo {title}
  {{Blueprint for a high-performance fluxonium quantum processor}},\ }\href
  {https://doi.org/10.1103/PRXQuantum.3.037001} {\bibfield  {journal} {\bibinfo
   {journal} {PRX Quantum}\ }\textbf {\bibinfo {volume} {3}},\ \bibinfo {pages}
  {037001} (\bibinfo {year} {2022})}\BibitemShut {NoStop}%
\bibitem [{\citenamefont {Rabi}(1936)}]{Rabi1936}%
  \BibitemOpen
  \bibfield  {author} {\bibinfo {author} {\bibfnamefont {I.~I.}\ \bibnamefont
  {Rabi}},\ }\bibfield  {title} {\bibinfo {title} {{On the process of space
  quantization}},\ }\href {https://doi.org/10.1103/PhysRev.49.324} {\bibfield
  {journal} {\bibinfo  {journal} {Phys. Rev.}\ }\textbf {\bibinfo {volume}
  {49}},\ \bibinfo {pages} {324} (\bibinfo {year} {1936})}\BibitemShut
  {NoStop}%
\bibitem [{\citenamefont {Rabi}(1937)}]{Rabi1937}%
  \BibitemOpen
  \bibfield  {author} {\bibinfo {author} {\bibfnamefont {I.~I.}\ \bibnamefont
  {Rabi}},\ }\bibfield  {title} {\bibinfo {title} {{Space quantization in a
  gyrating magnetic field}},\ }\href {https://doi.org/10.1103/PhysRev.51.652}
  {\bibfield  {journal} {\bibinfo  {journal} {Phys. Rev.}\ }\textbf {\bibinfo
  {volume} {51}},\ \bibinfo {pages} {652} (\bibinfo {year} {1937})}\BibitemShut
  {NoStop}%
\bibitem [{\citenamefont {Krantz}\ \emph {et~al.}(2019)\citenamefont {Krantz},
  \citenamefont {Kjaergaard}, \citenamefont {Yan}, \citenamefont {Orlando},
  \citenamefont {Gustavsson},\ and\ \citenamefont {Oliver}}]{Krantz2019}%
  \BibitemOpen
  \bibfield  {author} {\bibinfo {author} {\bibfnamefont {P.}~\bibnamefont
  {Krantz}}, \bibinfo {author} {\bibfnamefont {M.}~\bibnamefont {Kjaergaard}},
  \bibinfo {author} {\bibfnamefont {F.}~\bibnamefont {Yan}}, \bibinfo {author}
  {\bibfnamefont {T.~P.}\ \bibnamefont {Orlando}}, \bibinfo {author}
  {\bibfnamefont {S.}~\bibnamefont {Gustavsson}},\ and\ \bibinfo {author}
  {\bibfnamefont {W.~D.}\ \bibnamefont {Oliver}},\ }\bibfield  {title}
  {\bibinfo {title} {{A quantum engineer's guide to superconducting qubits}},\
  }\href {https://doi.org/10.1063/1.5089550} {\bibfield  {journal} {\bibinfo
  {journal} {Appl. Phys. Rev.}\ }\textbf {\bibinfo {volume} {6}},\ \bibinfo
  {pages} {021318} (\bibinfo {year} {2019})}\BibitemShut {NoStop}%
\bibitem [{\citenamefont {Rasmussen}\ \emph {et~al.}(2021)\citenamefont
  {Rasmussen}, \citenamefont {Christensen}, \citenamefont {Pedersen},
  \citenamefont {Kristensen}, \citenamefont {B{\ae}kkegaard}, \citenamefont
  {Loft},\ and\ \citenamefont {Zinner}}]{Rasmussen2021}%
  \BibitemOpen
  \bibfield  {author} {\bibinfo {author} {\bibfnamefont {S.~E.}\ \bibnamefont
  {Rasmussen}}, \bibinfo {author} {\bibfnamefont {K.~S.}\ \bibnamefont
  {Christensen}}, \bibinfo {author} {\bibfnamefont {S.~P.}\ \bibnamefont
  {Pedersen}}, \bibinfo {author} {\bibfnamefont {L.~B.}\ \bibnamefont
  {Kristensen}}, \bibinfo {author} {\bibfnamefont {T.}~\bibnamefont
  {B{\ae}kkegaard}}, \bibinfo {author} {\bibfnamefont {N.~J.}\ \bibnamefont
  {Loft}},\ and\ \bibinfo {author} {\bibfnamefont {N.~T.}\ \bibnamefont
  {Zinner}},\ }\bibfield  {title} {\bibinfo {title} {{Superconducting circuit
  companion---an introduction with worked examples}},\ }\href
  {https://doi.org/10.1103/PRXQuantum.2.040204} {\bibfield  {journal} {\bibinfo
   {journal} {PRX Quantum}\ }\textbf {\bibinfo {volume} {2}},\ \bibinfo {pages}
  {040204} (\bibinfo {year} {2021})}\BibitemShut {NoStop}%
\bibitem [{\citenamefont {Zeuthen}\ \emph {et~al.}(2018)\citenamefont
  {Zeuthen}, \citenamefont {Schliesser}, \citenamefont {Taylor},\ and\
  \citenamefont {S{\o}rensen}}]{Zeuthen2018}%
  \BibitemOpen
  \bibfield  {author} {\bibinfo {author} {\bibfnamefont {E.}~\bibnamefont
  {Zeuthen}}, \bibinfo {author} {\bibfnamefont {A.}~\bibnamefont {Schliesser}},
  \bibinfo {author} {\bibfnamefont {J.~M.}\ \bibnamefont {Taylor}},\ and\
  \bibinfo {author} {\bibfnamefont {A.~S.}\ \bibnamefont {S{\o}rensen}},\
  }\bibfield  {title} {\bibinfo {title} {{Electrooptomechanical equivalent
  circuits for quantum transduction}},\ }\href
  {https://doi.org/10.1103/PhysRevApplied.10.044036} {\bibfield  {journal}
  {\bibinfo  {journal} {Phys. Rev. Appl.}\ }\textbf {\bibinfo {volume} {10}},\
  \bibinfo {pages} {044036} (\bibinfo {year} {2018})}\BibitemShut {NoStop}%
\bibitem [{\citenamefont {Bl{\'{e}}sin}\ \emph {et~al.}(2021)\citenamefont
  {Bl{\'{e}}sin}, \citenamefont {Tian}, \citenamefont {Bhave},\ and\
  \citenamefont {Kippenberg}}]{Blesin2021}%
  \BibitemOpen
  \bibfield  {author} {\bibinfo {author} {\bibfnamefont {T.}~\bibnamefont
  {Bl{\'{e}}sin}}, \bibinfo {author} {\bibfnamefont {H.}~\bibnamefont {Tian}},
  \bibinfo {author} {\bibfnamefont {S.~A.}\ \bibnamefont {Bhave}},\ and\
  \bibinfo {author} {\bibfnamefont {T.~J.}\ \bibnamefont {Kippenberg}},\
  }\bibfield  {title} {\bibinfo {title} {{Quantum coherent microwave-optical
  transduction using high-overtone bulk acoustic resonances}},\ }\href
  {https://doi.org/10.1103/PhysRevA.104.052601} {\bibfield  {journal} {\bibinfo
   {journal} {Phys. Rev. A}\ }\textbf {\bibinfo {volume} {104}},\ \bibinfo
  {pages} {052601} (\bibinfo {year} {2021})}\BibitemShut {NoStop}%
\bibitem [{\citenamefont {Miyanaga}\ \emph {et~al.}(2021)\citenamefont
  {Miyanaga}, \citenamefont {Tomonaga}, \citenamefont {Ito}, \citenamefont
  {Mukai},\ and\ \citenamefont {Tsai}}]{Miyanaga2021}%
  \BibitemOpen
  \bibfield  {author} {\bibinfo {author} {\bibfnamefont {T.}~\bibnamefont
  {Miyanaga}}, \bibinfo {author} {\bibfnamefont {A.}~\bibnamefont {Tomonaga}},
  \bibinfo {author} {\bibfnamefont {H.}~\bibnamefont {Ito}}, \bibinfo {author}
  {\bibfnamefont {H.}~\bibnamefont {Mukai}},\ and\ \bibinfo {author}
  {\bibfnamefont {J.~S.}\ \bibnamefont {Tsai}},\ }\bibfield  {title} {\bibinfo
  {title} {{Ultrastrong tunable coupler between superconducting LC
  resonators}},\ }\href {https://doi.org/10.1103/PHYSREVAPPLIED.16.064041}
  {\bibfield  {journal} {\bibinfo  {journal} {Phys. Rev. Appl.}\ }\textbf
  {\bibinfo {volume} {16}},\ \bibinfo {pages} {064041} (\bibinfo {year}
  {2021})}\BibitemShut {NoStop}%
\bibitem [{\citenamefont {Prosen}(2008)}]{Prosen2008}%
  \BibitemOpen
  \bibfield  {author} {\bibinfo {author} {\bibfnamefont {T.}~\bibnamefont
  {Prosen}},\ }\bibfield  {title} {\bibinfo {title} {{Third quantization: A
  general method to solve master equations for quadratic open Fermi systems}},\
  }\href {https://doi.org/10.1088/1367-2630/10/4/043026} {\bibfield  {journal}
  {\bibinfo  {journal} {New J. Phys.}\ }\textbf {\bibinfo {volume} {10}},\
  \bibinfo {pages} {043026} (\bibinfo {year} {2008})}\BibitemShut {NoStop}%
\bibitem [{\citenamefont {McDonald}\ and\ \citenamefont
  {Clerk}(2023)}]{McDonald2023}%
  \BibitemOpen
  \bibfield  {author} {\bibinfo {author} {\bibfnamefont {A.}~\bibnamefont
  {McDonald}}\ and\ \bibinfo {author} {\bibfnamefont {A.~A.}\ \bibnamefont
  {Clerk}},\ }\bibfield  {title} {\bibinfo {title} {{Third quantization of open
  quantum systems: Dissipative symmetries and connections to phase-space and
  Keldysh field-theory formulations}},\ }\href
  {https://doi.org/10.1103/PhysRevResearch.5.033107} {\bibfield  {journal}
  {\bibinfo  {journal} {Phys. Rev. Res.}\ }\textbf {\bibinfo {volume} {5}},\
  \bibinfo {pages} {033107} (\bibinfo {year} {2023})}\BibitemShut {NoStop}%
\bibitem [{\citenamefont {Cahill}\ and\ \citenamefont
  {Glauber}(1969)}]{Cahill1969}%
  \BibitemOpen
  \bibfield  {author} {\bibinfo {author} {\bibfnamefont {K.~E.}\ \bibnamefont
  {Cahill}}\ and\ \bibinfo {author} {\bibfnamefont {R.~J.}\ \bibnamefont
  {Glauber}},\ }\bibfield  {title} {\bibinfo {title} {{Density operators and
  quasiprobability distributions}},\ }\href
  {https://doi.org/10.1103/PhysRev.177.1882} {\bibfield  {journal} {\bibinfo
  {journal} {Phys. Rev.}\ }\textbf {\bibinfo {volume} {177}},\ \bibinfo {pages}
  {1882} (\bibinfo {year} {1969})}\BibitemShut {NoStop}%
\bibitem [{\citenamefont {Angeletti}\ \emph {et~al.}(2023)\citenamefont
  {Angeletti}, \citenamefont {Zippilli},\ and\ \citenamefont
  {Vitali}}]{Angeletti2023}%
  \BibitemOpen
  \bibfield  {author} {\bibinfo {author} {\bibfnamefont {J.}~\bibnamefont
  {Angeletti}}, \bibinfo {author} {\bibfnamefont {S.}~\bibnamefont
  {Zippilli}},\ and\ \bibinfo {author} {\bibfnamefont {D.}~\bibnamefont
  {Vitali}},\ }\bibfield  {title} {\bibinfo {title} {{Dissipative stabilization
  of entangled qubit pairs in quantum arrays with a single localized
  dissipative channel}},\ }\href {https://doi.org/10.1088/2058-9565/acd4e3}
  {\bibfield  {journal} {\bibinfo  {journal} {Quantum Sci. Technol.}\ }\textbf
  {\bibinfo {volume} {8}},\ \bibinfo {pages} {035020} (\bibinfo {year}
  {2023})}\BibitemShut {NoStop}%
\bibitem [{\citenamefont {Johansson}\ \emph {et~al.}(2012)\citenamefont
  {Johansson}, \citenamefont {Nation},\ and\ \citenamefont
  {Nori}}]{Johansson2012}%
  \BibitemOpen
  \bibfield  {author} {\bibinfo {author} {\bibfnamefont {J.~R.}\ \bibnamefont
  {Johansson}}, \bibinfo {author} {\bibfnamefont {P.~D.}\ \bibnamefont
  {Nation}},\ and\ \bibinfo {author} {\bibfnamefont {F.}~\bibnamefont {Nori}},\
  }\bibfield  {title} {\bibinfo {title} {{QuTiP: An open-source Python
  framework for the dynamics of open quantum systems}},\ }\href
  {https://doi.org/10.1016/j.cpc.2012.02.021} {\bibfield  {journal} {\bibinfo
  {journal} {Comput. Phys. Commun.}\ }\textbf {\bibinfo {volume} {183}},\
  \bibinfo {pages} {1760} (\bibinfo {year} {2012})}\BibitemShut {NoStop}%
\bibitem [{\citenamefont {Johansson}\ \emph {et~al.}(2013)\citenamefont
  {Johansson}, \citenamefont {Nation},\ and\ \citenamefont
  {Nori}}]{Johansson2013}%
  \BibitemOpen
  \bibfield  {author} {\bibinfo {author} {\bibfnamefont {J.~R.}\ \bibnamefont
  {Johansson}}, \bibinfo {author} {\bibfnamefont {P.~D.}\ \bibnamefont
  {Nation}},\ and\ \bibinfo {author} {\bibfnamefont {F.}~\bibnamefont {Nori}},\
  }\bibfield  {title} {\bibinfo {title} {{QuTiP 2: A Python framework for the
  dynamics of open quantum systems}},\ }\href
  {https://doi.org/10.1016/j.cpc.2012.11.019} {\bibfield  {journal} {\bibinfo
  {journal} {Comput. Phys. Commun.}\ }\textbf {\bibinfo {volume} {184}},\
  \bibinfo {pages} {1234} (\bibinfo {year} {2013})}\BibitemShut {NoStop}%
\end{thebibliography}%

\end{document}